\documentclass[aps,pre,twocolumn,floatfix,longbibliography,superscriptaddress]{revtex4-1}
\pdfoutput=1
\usepackage{graphicx}
\usepackage{mathtools}
\usepackage{pgffor}
\usepackage{pdfpages} 
\usepackage{siunitx}
\usepackage{xcolor}
\usepackage[normalem]{ulem}
\usepackage{hyperref}
\definecolor{ngreen}{RGB}{0,153,0}
\definecolor{nred}{RGB}{153,0,0}
\makeatletter
\AtBeginDocument{\let\LS@rot\@undefined}
\makeatother

\begin{document}

\title{Resolving the degree of order in the bacterial \\ chromosome using a statistical physics approach}

\author{Joris J.B. Messelink}
\affiliation{Arnold Sommerfeld Center for Theoretical Physics and Center for NanoScience, Department of Physics, Ludwig Maximilian-University Munich, Theresienstr. 37, D-80333 Munich, Germany}
\author{Jacqueline Janssen}
\affiliation{Arnold Sommerfeld Center for Theoretical Physics and Center for NanoScience, Department of Physics, Ludwig Maximilian-University Munich, Theresienstr. 37, D-80333 Munich, Germany}
\author{Muriel C.F. van Teeseling}
\affiliation{Department of Biology, University of Marburg, Germany}
\author{Martin Thanbichler}
\affiliation{Department of Biology, University of Marburg, Germany}
\affiliation{Max Planck Institute for Terrestrial Microbiology, Marburg, Germany}
\affiliation{Center for Synthetic Microbiology, Marburg, Germany}
\author{Chase P. Broedersz}
\affiliation{Arnold Sommerfeld Center for Theoretical Physics and Center for NanoScience, Department of Physics, Ludwig Maximilian-University Munich, Theresienstr. 37, D-80333 Munich, Germany}
\affiliation{Correspondence: c.broedersz@lmu.de}

\begin{abstract}
\noindent While bacterial chromosomes were long thought to be amorphous, recent experiments reveal pronounced organizational features. However, the extent of bacterial chromosome organization remains unclear. Here, we develop a fully data-driven maximum entropy approach to extract the distribution of single-cell chromosome conformations from experimental normalized Hi-C data. We apply this inference to the model organism \textit{Caulobacter crescentus}. On small genomic scales of $10^4$-$10^5$ basepairs, our model reveals a pattern of local chromosome extensions that correlates with transcriptional and DNA loop extrusion activity. On larger genomic scales, we find that chromosome structure is predominantly present along the long cell axis: chromosomal loci not only have well-defined axial positions, they also exhibit long-ranged correlations due interacting large emergent genomic clusters, termed Super Domains. Finally, our model reveals  information contained in chromosome structure that can guide cellular processes.  Our approach can be generalized to other species, providing a principled way of analyzing spatial chromosome organization. 
\end{abstract}

\maketitle

\noindent Chromosomes carry all information to generate a living cell. In many bacteria this information is stored on a single circular chromosome, with a length three orders of magnitude larger than the cell. This implies a major organizational problem~\cite{Thanbichler2006,wang,Badrinarayanan2015}: The DNA not only has to be condensed to fit in the bacterial cell,  its organization also needs to facilitate functions such as transcription and replication. Various proteins regulate chromosome structure~\cite{Dame2019,Dillon2010,Broedersz2014,Graham2014,Brackley2017}, but it remains unclear how organized it is across all length scales. Resolving this organization requires a characterization of the distribution of single-cell chromosome conformations, posing a key challenge for experiment and theory~\cite{Imakaev2015}.

The classical picture in which the bacterial chromosome is arranged as an amorphous polymer has become obsolete thanks to recent experimental advances~\cite{Robinett1996,Cattoni2015,Wu2019,lieberman}. Indeed,  fluorescence microscopy experiments revealed that chromosomal loci localize to well-defined cellular addresses in various species~\cite{Teleman1998,Wiggins2010,Bates2005,Lau2004}, including \textit{Caulobacter crescentus}~\cite{Viollier2004}. Further insights were obtained by chromosome conformation capture 5C/Hi-C experiments~\cite{Umbarger2012,Le2014}, measuring average pair-wise contacts between  loci. These experiments revealed Chromosomal Interaction Domains (CIDs) of up to $10^5$ basepairs, comprising loci preferentially interacting within their domain. Various processes~\cite{Marbouty2015,Lioy2018},  including transcription~\cite{Le2016,le}, impact CID organization. On larger genomic scales, locus pairs on opposite chromosomal arms often favor a juxtaposed arrangement, induced by the loop extrusion motor SMC~\cite{le,umbarger,tran2017,Marbouty2015,Wang2015,Wang2017,Bohm2019}. Despite these observations, the degree of structural order in the bacterial chromosome still remains elusive.

To exploit advances in Hi-C experiments on various bacteria~\cite{le,Marbouty2015,Wang2015,Lioy2018,Bohm2019, Trussart2017}, a principled data-driven theoretical  approach is needed. However,  there are several outstanding challenges that preclude a fully data-driven model~\cite{le,umbarger,Yildirim2018,Imakaev2015}. Hi-C data is typically normalized, and it is unclear how to relate normalized Hi-C scores to theoretical contact frequencies. Previous approaches~\cite{umbarger,Yildirim2018,Oluwadare2019} rely on an assumed relation  between Hi-C scores and the average spatial distance between locus pairs, and this constraint is independently enforced on  each pair, ignoring correlations. Alternative methods  generate configuration ensembles, e.g.  using iterative maximum likelihood algorithms~\cite{Tjong2016}. However, Hi-C maps could be consistent with many ensembles. Thus, a principled criterium is needed to select an unbiased configuration distribution with high predictive power. For eukaryotes, an equilibrium Maximum Entropy (MaxEnt) selection method was proposed~\cite{Zhang2015,DiPierro2016,Abbas2019}, as used for protein structure prediction~\cite{Weigt2009,Marks2011}. However, such an approach may be unsuitable for chromosomes in living cells, which exhibit non-equilibrium fluctuations~\cite{Weber2012,Javer2013,Smith2015}. Thus, a principled and unbiased approach to derive a unique non-equilibrium model for the distribution of chromosome conformations is still lacking.

Here, we develop a \textit{fully} data-driven MaxEnt approach for the bacterial chromosome based on Hi-C data. This approach infers the least-structured distribution of chromosome conformations that fits Hi-C experiments, capturing population heterogeneity at the single-cell level.  Our MaxEnt model  does not rely on equilibrium assumptions, it is inferred directly from normalized Hi-C scores, it does not require an assumed Hi-C---distance relation, and we determine the coarse-graining scale of our model using experiments. The MaxEnt model reveals structural features over a broad range of genomic length scales, and we quantify the positional information in the cellular location of chromosomal loci that can be used by cellular processes. Our theoretical framework may be generalized to other prokaryotic and eukaryotic species, providing a rigorous approach to resolve chromosome organization from Hi-C data. 

\section{Maximum entropy model inferred from chromosomal contact frequencies}
\label{sec:model} 
\noindent Our goal is to determine the ensemble of single-cell chromosome conformations for a heterogeneous cell population from experimental Hi-C data. To this end, we build on existing MaxEnt methods for analysing biophysical data~\cite{Tkacik2013,Zhang2015,DiPierro2016,Marks2011,Mora2010,Bialek2012}, to develop a principled approach for inferring the statistics of chromosome structure in bacteria from experiments.

The microstates $\{ \sigma \}$ of the system are defined as the set of all  configurations of the chromosome contained within the cellular confinement. We seek the statistical weights $P(\sigma)$, chosen to be consistent with the experimental Hi-C  map. In general, however, a set of experimental constraints does not uniquely determine $P(\sigma)$. The MaxEnt approach is based on selecting $P(\sigma)$ from these possible solutions by choosing the unique distribution with the largest Shannon entropy, 
\begin{equation}\label{eq:shannon}
S = - \sum_{\sigma} P(\sigma) \ln P(\sigma),
\end{equation}
constituting the least-structured distribution consistent with experimental data. Put simply, we require that the only structure present in $P(\sigma)$ is due to experimental constraints from Hi-C scores, rather than assumed features of the underlying polymer model, the interpretation of Hi-C scores, or the ensemble-generating algorithm.

To apply the MaxEnt method to experimental Hi-C data, we employ a coarse-grained representation of the chromosome: a polymer on a 3D cubic lattice, with a subset of monomers representing $N$ genomic regions. This provides an efficient computational framework, while still capturing key organizational features. Specifically, this representation is chosen to preserve experimentally measured distance fluctuations at the coarse-graining scale (Methods and SI S1-2). At larger scales, the statistics of polymer configurations are only constrained by Hi-C data. Within this representation, a microstate $\sigma=\{{\bf r}_1, {\bf r}_2, ...\}=\{{\bf r}\}$ is defined by the monomer positions ${\bf r}_i$. Two genomic regions have a contact probability $\gamma$ if they occupy the same lattice site, and $0$ otherwise. 

To obtain the least-structured distribution of configurations consistent with experiments, we seek $P(\{{\bf r}\})$ that maximizes $S$ (Eq.~\eqref{eq:shannon}) under two constraints: 1) the model contact frequencies should match experimental contact frequencies $f^{\rm expt}_{ij}$ between genomic regions $i$ and $j$ (the correspondence between $f^{\rm expt}_{ij}$ and Hi-C scores is discussed in the next section), and 2) the distribution should be normalized. To this end, we introduce  constraints to the entropy functional:
\begin{align}\label{eq:EntropyFunctional}
\tilde{S} = &- \sum_{\{{\bf r}\}} P(\{{\bf r}\}) \ln P(\{{\bf r}\}) - \sum_{ij} \lambda_{ij} \biggl(  \sum_{\{{\bf r}\}} P(\{{\bf r}\}) \gamma \delta_{{\bf r}_i,{\bf r}_j}   \nonumber \\
&- f^{{\rm expt}}_{ij} \biggr)- \lambda_0 \biggl(\sum_{\{{\bf r}\}} P(\{{\bf r}\})-1\biggr)
\end{align}
For each  data point $f_{ij}^{\rm expt}$, there is a corresponding Lagrange multiplier $\lambda_{ij}$, and $\delta_{{\bf r}_i,{\bf r}_j}$ is the Kronecker delta. In addition, the Lagrange multiplier $\lambda_0$ ensures normalization.  We maximize the Shannon entropy under these constraints, setting $\tfrac{\delta \tilde{S}}{\delta P(\{{\bf r}\})} = 0$, yielding
\begin{equation} 
\label{eq:pofsigma}
P(\{{\bf r}\}) = \frac{1}{Z} \exp\left[ -\sum_{ij}\lambda_{ij} \gamma \delta_{{\bf r}_i,{\bf r}_j} \right],
\end{equation}
with $Z = \exp[1+\lambda_0]$. The $\lambda_{ij}$'s paramatrizing $P(\{{\bf r}\})$ are determined by solving
\begin{equation}
 \sum_{\{{\bf r}\}} P(\{{\bf r}\}) \gamma \delta_{{\bf r}_i,{\bf r}_j} = f^{{\rm expt}}_{ij}
 \end{equation}
 for each experimental constraint. For typical Hi-C data on a bacterial chromosome, this amounts to of order $10^5$ constraints~\cite{le}. These equations can not be solved directly, as they are highly nonlinear and the state space is very large.
 
 The daunting challenge of finding the Lagrange multipliers can be overcome by noting that the distribution in Eq.~\eqref{eq:pofsigma} can be mapped to a statistical mechanics model: a confined lattice polymer, with a (dimensionless) Hamiltonian
 \begin{equation}\label{eq:Heff}
 H = \frac{1}{2}\sum_{ij} \epsilon_{ij} \delta_{{\bf r}_i,{\bf r}_j}, 
 \end{equation}
where  $\epsilon_{ij}= \gamma \lambda_{ij}$ are the effective interaction energies between overlapping loci. We numerically obtain the inverse solutions of this  model using iterative Monte Carlo simulations (SI S3). Testing this algorithm on contact frequency maps generated from a set of chosen input $\epsilon_{ij}$, we find that our algorithm precisely and robustly recovers the correct input values (SI S4).

\section{Inferring the MaxEnt model directly from normalized Hi-C scores}\label{sec:conversion}
\noindent A major hurdle in applying data-driven inference approaches is finding a correspondence between experimental Hi-C scores and the contact frequencies in a coarse-grained polymer model. Published Hi-C maps are typically normalized~\cite{Pal2019}. This normalization compensates known biases in raw Hi-C data, for instance due to the proportionality between the number of restriction sites in a genomic region and its Hi-C score. Furthermore, absolute Hi-C scores are hard to interpret because it is difficult to estimate the conversion factor to physical contact frequencies. Importantly however, even if absolute contact scores could be obtained, a mapping to contact frequencies in a coarse-grained model is challenging.

We address this conversion issue by treating the conversion factor as an unknown parameter $c$ in our MaxEnt procedure. Thus, we write $f_{ij}^{\rm expt} = c \tilde{f}_{ij}^{\rm expt}$, with  $\tilde{f}_{ij}^{\rm expt}$ the normalized experimental Hi-C scores. We  absorb the contact probability factor $\gamma$ into $c$ (Eq.~\eqref{eq:EntropyFunctional}), setting  $\tilde{c} = c \gamma$, and require that $\tilde{c}$ maximizes the model entropy (SI S3.2), yielding the additional constraint 
\begin{equation}
\label{eq:conditionc}
\sum_{ij} \epsilon_{ij}  \tilde{f}_{ij}^{\rm expt} = 0.
\end{equation}  
Thus, we infer the least-structured distribution of chromosome conformations from normalized Hi-C data, without assuming a conversion  between Hi-C scores and contact frequencies or average distances between loci.

\begin{figure}[t!]
  \includegraphics[width = \columnwidth]{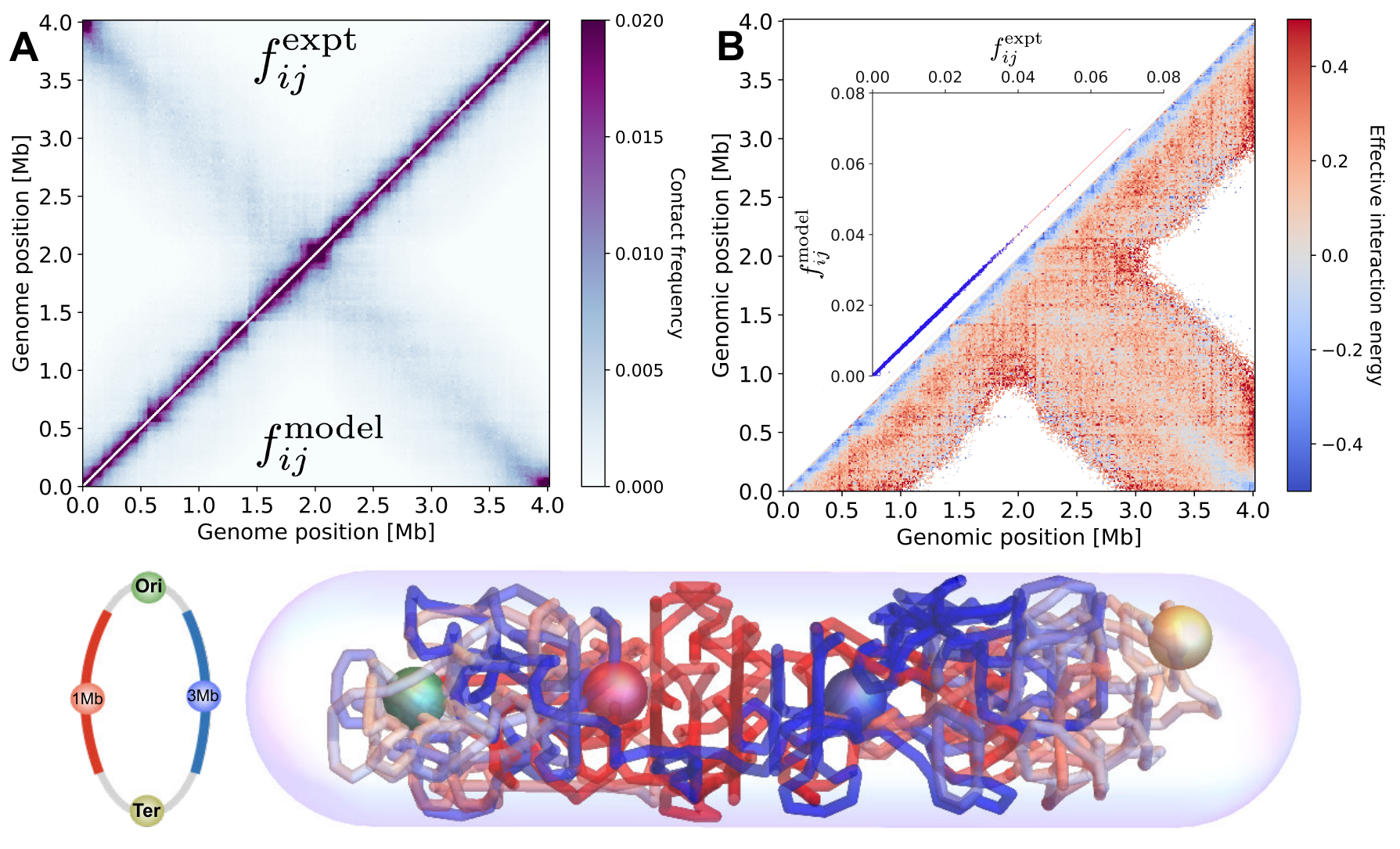}
  \caption{\label{fig:expt_compare} \textbf{Maximum entropy model inferred from Hi-C experiments in \textit{C. crescentus}.} \textbf{A} Comparison between experimental contact frequencies $f_{ij}^{\rm expt}$ (upper left corner, adapted from Ref.~\cite{le}) and contact frequencies obtained from our inferred MaxEnt model  $f_{ij}^{\rm model}$ (lower right corner). \textbf{B} Inferred effective interaction energies $\epsilon_{ij}$ (lower right corner, white regions indicate $\epsilon_{ij}\rightarrow\infty$) together with scatter plot of $f_{ij}^{\rm expt}$ vs. $f_{ij}^{\rm model}$ (inset). \textbf{C} Visualization of a single-cell chromosome configuration predicted by our MaxEnt model; the centers of four distinct chromosome sections are represented  in the schematic by colored spheres.}
\end{figure}

\section{MaxEnt model of the \textit{C. crescentus} chromosome quantitatively captures measured cellular localization}
We investigate the degree of organization of the bacterial chromosome by considering newborn swarmer cells of the model organism \textit{C. crescentus}. To develop the MaxEnt model, we first experimentally determine the coarse-graining scale, set by the average distance between consecutive 10 kb genomic regions (Methods SI 1-2). Subsequently, we infer the parameters of the MaxEnt model from published experimental Hi-C data (SI S5)~\cite{le}. Our inverse algorithm robustly converges to an accurate description of the Hi-C map: the modelled and experimental contact map agree within $3.1\%$ with a Pearson's correlation coefficient of 0.9996 (Fig.~\ref{fig:expt_compare}A, B inset). 

Our MaxEnt model quantitatively reproduces essential features of the experimental Hi-C map (Fig.~\ref{fig:expt_compare}A), including the fine structure of the CIDs as well as the secondary diagonal, which is attributed to  the loop extrusion activity of SMC (Structural Maintenance of the Chromosome)\cite{Wang2017,Burmann2015,Miermans2018,Ganji2018}.
The inferred $\epsilon_{ij}$'s (Fig.~\ref{fig:expt_compare}B) should not be interpreted as physical interaction energies. Rather, they parametrize the predicted physical distribution of chromosome configurations $P(\{{\bf r}_i\})$. We can directly interpret the organizational features implied by $P(\{{\bf r}_i\})$ and use it to sample single-cell configurations (Fig.~\ref{fig:expt_compare}C).

\begin{figure}[t!]
  \includegraphics[width = \columnwidth]{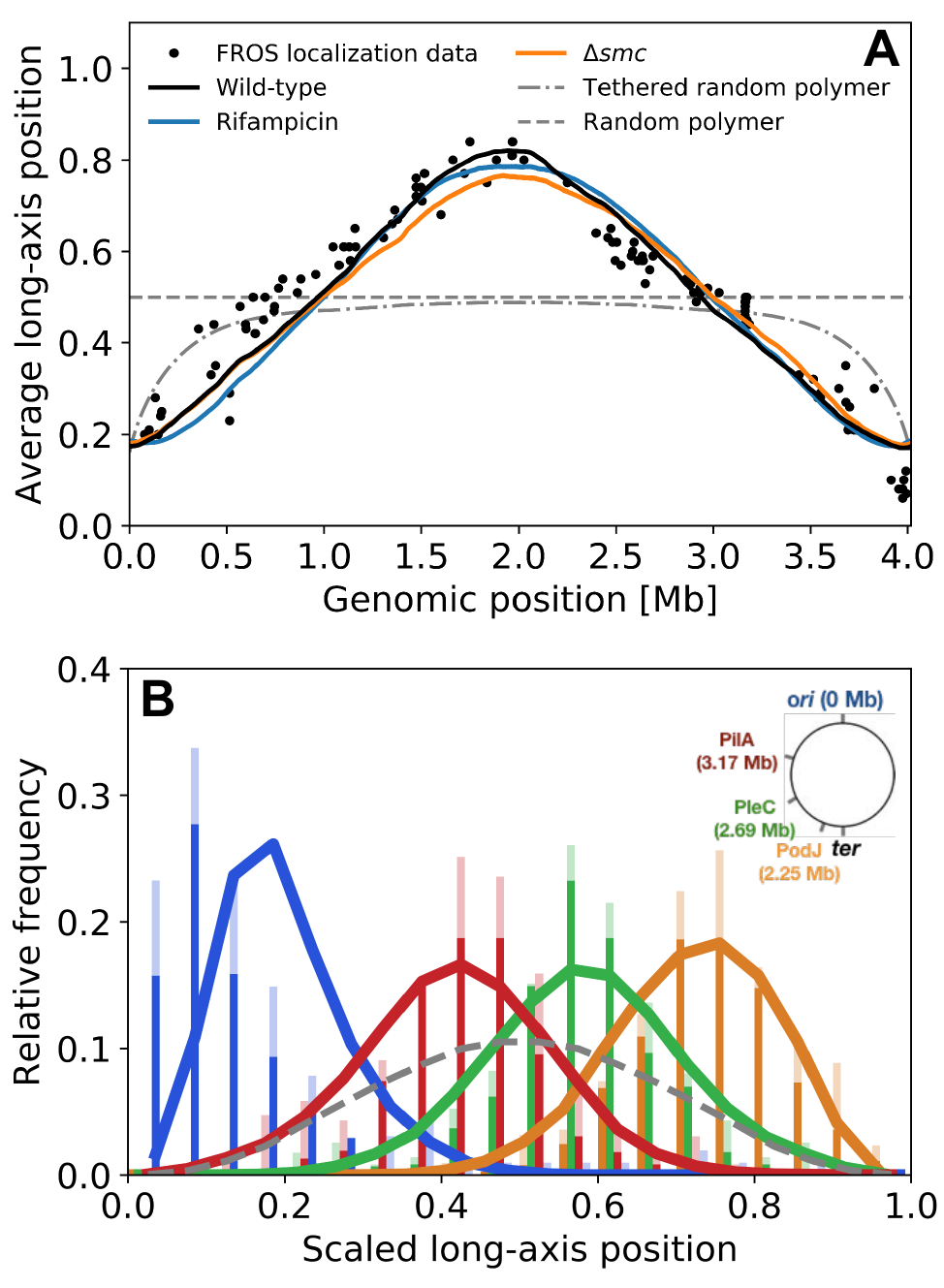}
  \caption{ \label{fig:localization_compare} \textbf{Validation of MaxEnt model based on spatial location microscopy data} \textbf{A} Average scaled long-axis position predicted from MaxEnt models (solid lines) inferred from various Hi-C data sets (from~\cite{le}), including wild-type cells (black), rifampicin-treated cells (blue), and $\Delta$\textit{smc} cells (orange), together with results from microscopy experiments (adapted from~\cite{Viollier2004}). Also shown are simulated data for a random polymer with \textit{ori}-pole tether (dash-dotted grey line), and a simulated confined random polymer (dashed grey line). \textbf{B} 
 The single cell distribution  of chromosomal loci (color coded) inside the cell (scaled long-axis position) as predicted by the MaxEnt model (solid lines), fits previous experimental data from microscopy experiments (bars, adapted from~\cite{Viollier2004}). To indicate experimental variability, the solid/transparent bars indicate the minimum/maximum measured by two different methods: FROS or FISH. The dashed grey line indicates the distribution for a confined random polymer.}
\end{figure}

We test the predictive power of the MaxEnt model by computing the distribution of axial locations of several loci. Importantly, we do not assume (polar) cell envelope tethering of specific loci, such as the origin of replication (\textit{ori}). We orient  cells by setting the \textit{ori} pole in the cell-half containing \textit{ori}. Interestingly, we find a high degree of axial localization of loci: the average axial position of loci is roughly linearly organized, and the predicted positions match previous live-cell microscopy experiments~\cite{Viollier2004} (Fig.~\ref{fig:localization_compare}A). By contrast, a confined random polymer---not constrained by Hi-C data---does not exhibit the linear organization, even when \textit{ori} is  tethered to the cell pole.

The MaxEnt model also predicts distributions of long-axis positions of chromosomal loci, in remarkable agreement with prior experiments (Fig.~\ref{fig:localization_compare}B). This comparison with independent experimental data constitutes a strong validation of our MaxEnt model.  The slight deviation of  position of \textit{ori} compared to the experiment (Fig.~\ref{fig:localization_compare}A,B) can be addressed with an extended MaxEnt model that incorporates the distribution of axial \textit{ori} positions as an additional constraint (SI S12). However, other aspects of the predicted chromosomal organization are largely unaffected by this modification, and therefore we will not impose this additional constraint in our analysis.

\section{Large-scale chromosome organization primarily characterized by long-axis correlations due to Super Domains} 
\begin{figure}[t!]
\includegraphics[width = \columnwidth]{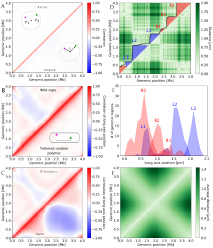}
\caption{ \label{fig:Lproperties} \textbf{MaxEnt model predicts large-scale features of chromosome organization} \textbf{A} Upper left corner: two-point correlations in the radial positions between genomic regions. Lower right corner: two-point correlations in angular orientations around the long axis. \textbf{B} Upper left corner: two-point correlations between long-axis positions for wild-type cells. Lower right corner: the same correlations for a model not constrained by Hi-C data, but with a tethered origin (a tethered random polymer). \textbf{C} Correlations in the long-axis positions of genomic regions derived from Hi-C data~\cite{le} of two modified conditions: cells treated with rifampicinn to block transcription (upper left corner), and $\Delta$\textit{smc} cells (lower right corner).  \textbf{D} Distance map for pairs of genomic regions for one chromosomal configuration. The inferred outlines of Super Domains (SuDs)  are indicated by a black line, with left/right-arm SuDs shaded blue/red. \textbf{E} Long axis distribution of genomic regions in SuDs identified in the configuration depicted in D. \textbf{F} Average spatial distances between genomic regions.}
\end{figure}

\noindent Large-scale organizational features of the chromosome can be revealed by measuring various two-point correlation functions. Earlier models suggested a three-dimensional organization in which the two chromosomal arms wind around each other with roughly one helical turn~\cite{umbarger,Yildirim2018}. To test if this organization also emerges in our MaxEnt model, we compute two-point correlations of angular orientations. For each chromosome segment, we assign an orientation vector in the plane perpendicular to the long axis. We find that angular correlations  decay rapidly for genomic distances $\gtrsim0.2$Mb (Fig.~\ref{fig:Lproperties}A lower right).  Long-range negative correlations between the two chromosomal arms are thus negligible, indicating that a pronounced helical organization is not required to model the experimental Hi-C map.

The two-point correlation function in radial positions decays even more rapidly with genomic distance up to $\sim0.1$Mb (Fig.~\ref{fig:Lproperties}A upper left), indicating the absence of large-scale organization in this direction. By contrast, two-point correlations in the long-axis position exhibit a striking structure: we observe positive long-ranged correlations for pairs of  genomic regions on the same chromosomal arm, whereas correlations in axial positions between arms are predominantly negative (Fig.~\ref{fig:Lproperties}B upper left). These long-ranged correlations signify collective behavior. Importantly, for a model with a tethered origin not constrained by Hi-C data, such organization is absent (Fig.~\ref{fig:Lproperties}B lower right). 

We find that these intra-arm anticorrelations result from the spatial exclusion of large genomic clusters between the two chromosomal arms, which we term Super Domains (SuDs). SuDs emerge from a clustering analysis  of genomic regions (SI S6). The formation of domain-like structures is revealed by plotting the distance between pairs of loci for a specific chromosome configuration, with domains spanning up to a quarter of chromosome length (Fig.~\ref{fig:Lproperties}D-E). On average, SuDs contain $\sim60$ genomic regions; compared to CIDs, they are typically larger with a more variable size and genomic location across chromosome conformations. The variable and delocalized nature of SuDs is apparent from the average distance map between genomic regions, indicating no discrete structure (Fig.~\ref{fig:Lproperties}F). Importantly, SuDs forming on opposing chromosomal arms tend to spatially exclude each other (Fig.~\ref{fig:Lproperties}E): the fraction of overlap in axial positions is reduced by $19\%$ compared to randomly paired left and right arm configurations. This exclusion behavior translates to intra-arm anticorrelations for pairs of genomic regions with similar average axial positions.

To investigate the influence of cellular processes on long-axis organization, we perform the same analysis (SI S9) on published Hi-C data of rifampicin-treated cells and a mutant lacking SMC ($\Delta$\textit{smc})~\cite{le}. Rifampicin treatment inhibits transcription, whereas SMC actively extrudes DNA loops and induces juxtaposed chromosomal arms~\cite{wang,Burmann2015}. For both cases, our models predict an average localization along the long axis similar to the wild-type   (Fig.~\ref{fig:localization_compare}A). However, the predicted long-axis  correlations exhibit marked differences: for rifampicin-treated cells with inhibited transcription, anticorrelations between chromosomal arms are less pronounced (Fig.~\ref{fig:Lproperties}C upper left). In contrast, $\Delta$\textit{smc} cells display a  broad regime with strong anticorrelations between loci on opposite arms (Fig.~\ref{fig:Lproperties}C lower right). These effects are reflected in the statistics of SuDs: upon inhibition of transcription, the SuDs contain 10\% more genomic regions per domain than in the wild-type. Despite this increased density, the transcription-inhibited cells show a similar overlap of SuDs (16\% lower than for randomly paired arms). By contrast, $\Delta$\textit{smc} cells exhibit the same average SuD density as the wild-type, but a strong reduction of inter-arm domain overlap ($35 \%$ lower than for randomly paired arms). Thus, the action of SMC enhanced interactions between SuDs, whereas transcription alters their density.

\section{Local chromosome extension coincides with high transcription, but only for one chromosomal arm} 
\noindent The MaxEnt model provides access to local structural features that may be difficult to determine experimentally. Specifically, we consider the local chromosomal extension $\delta_i$, defined as the average spatial distance between  two neighbouring genomic regions of region $i$ (SI S11). Interestingly, the $\delta_i$-profile exhibits an overall trend that is lowest at \textit{ori} and \textit{ter} (Fig.~\ref{fig:Sproperties}A), indicating that these regions are intrinsically more compact (SI S11). In addition, pronounced peaks and valleys in local extension are revealed at a smaller genomic scale similar to that of CIDs. The same structure appears for $\Delta$\textit{smc} cells, although their chromosome appears to be locally more compact than the wild-type. By contrast, in rifampicin-cells, peak amplitudes are significantly suppressed, suggesting a link with transcription.

Previous work reported a connection between CID boundaries and highly transcribed genes~\cite{le}. Based on this observation and polymer simulations, it was suggested that high transcription creates plectoneme-free regions, physically separating CIDs. To further investigate the impact of gene expression activity on local structure, we compare the locations of local chromosome extension peaks in our MaxEnt model and the 2\% most highly transcribed genes. Indeed, we observe a  significantly increased overlap of local chromosome extension peaks and the locations of highly transcribed genes, compared to a random distribution of peaks, but only for genes on the forward strand of the right \textit{ori}-\textit{ter} arm (0-2.0 Mb) (SI S7). If the colocalization of local extension peaks by highly transcribed genes would only depend on the relative direction of transcription and replication, this should also occur for highly transcribed genes on backward strands on the left arm, which we do not observe. Thus a feature is required to break this symmetry. While our results indicate a connection between high local chromosome extension and the direction of replication and transcription of highly transcribed genes, the underlying molecular mechanism is still unclear.

\section{Chromosomal structure provides positional information in the cell} 
\noindent The inferred structural features of the chromosome not only yield insights into cellular organization, they may also have functional significance: organizational features provide information that could guide cellular processes. For example, proteins with a high relative affinity to certain genomic regions will be positioned more precisely within the cell. 
In addition, this information may enable a mechanism to position protein droplets~\cite{Shin2017}, by nucleating on specific chromosomal regions, as  e.g. suggested for clusters of DNA-binding chromosome partitioning proteins~\cite{Broedersz2014}.

\begin{figure}[t!]
\includegraphics[width = \columnwidth]{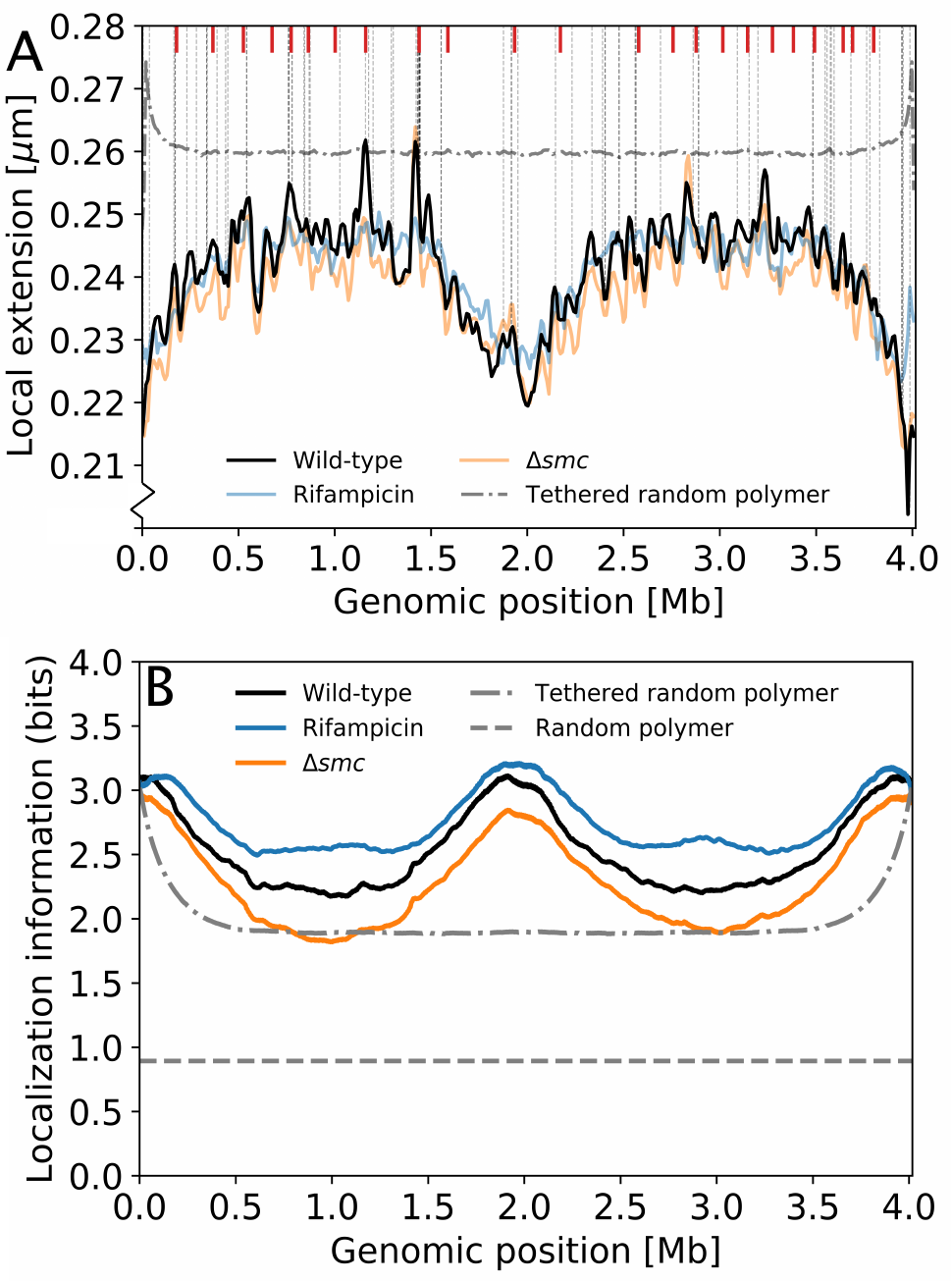}
\caption{ \label{fig:Sproperties} \textbf{MaxEnt model reveals local features and  positional information encoded by chromosome organization} \textbf{A} The local chromosome extension $\delta_i$ as a function of genomic position. $\delta_i$ is defined as the spatial distance between neighbouring genomic regions of site $i$ averaged over all chromosome conformations. Model prediction are shown for wild-type cells (black), rifampicin-treated cells (blue), $\Delta$\textit{smc}-cells (orange), and a pole-tethered random polymer (grey dash-dotted line). The locations of the top 2\% highly-transcribed genes are indicated by vertical grey dashed lines, the locations of CIDs determined in~\cite{le} are indicated by red markers.  \textbf{B} Positional information per genomic region in bits for wild-type (black), $\Delta$\textit{smc} (orange), rifampicin-treated cells (blue), a random pole-tethered polymer (dash-dotted line), and a random polymer (dashed line).}
\end{figure}

Using our MaxEnt model, we can quantify how much localization information (SI S10)~\cite{Dubuis2013} is encoded by chromosome organization per genomic regions (Fig.~\ref{fig:Sproperties}B). The information is largest near \textit{ori} and \textit{ter}, providing 3 bits of localization information, equivalent to reducing the positional uncertainty to one cellular octant. Comparing these results with those for modified conditions, we find that rifampicin treatment increases localization information, whereas information is reduced in $\Delta$\textit{smc} cells, suggesting that SMC action and transcription have opposing effects on localization information. This localization information is just one example of how structural features in the organization of the chromosome can be used to guide cellular processes. The MaxEnt approach provides a scheme to estimate the information available to the cell that is contained in the distribution of chromosome conformations.

\section{Discussion}
\noindent We established a fully data-driven principled approach to infer the spatial organization of the bacterial chromosome at the single-cell level, and applied this approach to normalized Hi-C data of the model organism \textit{C. crescentus}. The predictive power of this MaxEnt model is confirmed by prior microscopy experiments~\cite{Viollier2004} showing the distributions of axial positions of chromosomal loci within the cell. This approach could, however, also be extended towards an integrated MaxEnt model, simultaneously constrained by both Hi-C and such microscopy data~(SI S12). Contrary to previous modelling approaches, our MaxEnt model does not rely on an assumed connection between Hi-C scores and average spatial distances~\cite{Umbarger2012}. Instead, we can predict how these quantities are related: we find an approximate linear trend between intra-arm genomic distance and spatial distance (SI S8). However, there are substantial deviations from this trend, together with significant correlations in distances between genomic regions. Previous approaches could not account for such deviations and correlations. This may explain differences in model predictions such as the  helical structure suggested in~\cite{umbarger,Yildirim2018}, which we do not observe. 

By design, the MaxEnt model yields the least-structured distribution of chromosome conformations consistent with Hi-C experiments, allowing us to investigate the degree of order in the bacterial chromosome. To extract structural information from the MaxEnt model, we considered two-point correlation functions in the cellular positions of genomic regions. While we observe negligible radial and angular correlations, there are pronounced long-ranged correlations along the long cell axi, indicating collective behavior. This structure is related to the observation of variable and delocalized clusters of genomic regions, which we term Super Domains (SuDs). These SuDs might reflect blob-like structures observed with microscopy in \textit{Bacillus subtilis}~\cite{Marbouty2015} and \textit{Escherichia coli}~\cite{Wu2019}. 
Our MaxEnt model indicates a spatial exclusion of opposing SuDs from different chromosomal arms, resulting in longe-ranged anticorrelations in axial positions. Transcription and SMC have opposing effects on SuD properties: inter-arm overlap between domains is reduced by transcription and increased by SMC, consistent with the idea that SMC links chromosomal arms~\cite{Marbouty2015,Wang2015,Wang2017,Burmann2015}. At the smaller genomic scale of CIDs, we observe a characteristic pattern of local chromosomal extensions, being most compact at \textit{ori} and \textit{ter}. The compaction of the \textit{ori} region  may be due to the ParAB\textit{S} chromosome partitioning system~\cite{Broedersz2014,Graham2014}. However, it is still unclear if \textit{C. crescentus} contains other nucleoid-associated proteins~\cite{Dame2019,Dillon2010} that are involved in the compaction of other chromosomal regions such as \textit{ter}. Interestingly, peaks in local extension tend to coincide with highly transcribed genes, but only for the forward strand of the right chromosomal arm (SI S7). 

Using our MaxEnt model we estimated the cellular location information per genomic region. This information reaches up to 3 bits around \textit{ori} and \textit{ter},  equivalent to a positional uncertainty in the cell of one cellular octant. We speculate that such positional information encoded by the organization of the chromosome could be exploited for sub-cellular positioning of proteins and protein droplets. Our approach may be extended to other prokaryotes as well as eukaryotes, paving the road for unraveling an unprecedented amount of information on chromosome conformations at multiple length scales, elucidating single-cell variability and population averages.

\section{Acknowledgement}
{We thank Ben Machta for inspiring discussions, Karsten Miermans and Lucas Tröger for valuable input for the simulations, and Maritha Lippmann for excellent technical assistance. This research was funded by the Deutsche Forschungsgemeinschaft (DFG, German Research Foundation, Project 269423233 - TRR174). J.M. is supported by a DFG fellowship within the Graduate School of Quantitative Biosciences Munich (QBM).}

\section{Methods}
Here, we consider Hi-C data on \textit{C. crescentus} newborn swarmer cells~\cite{le}, which have a single, non-replicating chromosome.
Our algorithm (SI 3,4) requires two length scales: the dimensions of the cellular
confinement and the lattice spacing. As a cellular confinement, we use a
cylinder capped with hemispheres with the dimensions of a
newborn swarmer cell minus the cell envelope: $0.63 \mu m \times 2.2
\mu m$ (SI S1-2). A more detailed representation of the cellular
confinement shape does not appear to affect our main results (SI S12).
To set the coarse-graining scale of our MaxEnt model, we experimentally determined the distribution of spatial distances between subsequent Hi-C bins.
Specifically, the lattice spacing, $b$, is set by the
average spatial distance between consecutive 10kb regions  (the Hi-C bin size).
To determine this parameter, we probed the physical distance of two loci separated by 10kb in five different regions of the chromosome, using an approach comparable to~\cite{Hensel2013,Gaal2016}. To this end, we constructed strains whose chromosomes contained two independent arrays of transcription factor binding sites (comprising 10 LacI or TetR binding sites, respectively) inserted at the proper distance (SI S1). The sub-cellular positions of these arrays were then visualized by producing the respective fluorescently labeled transcription factors (LacI-eCFP and TetR-eYFP) at very low levels, based solely on the basal activity of the inducible promoter driving their expression. Swarmer cells were imaged immediately after isolation, and the localization of the two arrays
was determined with sub-pixel precision by fitting a 2D Gaussian to the acquired images.
The Euclidean distances between the two arrays were calculated,
taking into account correction factors for a systematic shift produced by the set-up (see
Methods for further details) and are shown in (Table S5).
The average distance between genomic loci 10 kb apart was found to be
$129 \pm 7$~nm, implying a lattice spacing $b=88$~nm (SI S2).

\bibliography{MaxEntCaulobacter4}



\section*{Supplementary material}

# Supplementary Information: Extracting the degree of order in the bacterial chromosome using a statistical physics approach

Joris J. B. Messelink[1], Jacqueline Janssen[1], Muriel C. F.van Teeseling[2], Martin Thanbichler[2,3,4], and Chase P. Broedersz[1,7]

[1]Arnold-Sommerfeld Center for Theoretical Physics and Center for NanoScience, Department of Physics, Ludwig Maximilian University of Munich, Theresienstr. 37, D-80333 Munich, Germany
[2]Department of Biology, University of Marburg, Germany
[3]Max Planck Institute for Terrestrial Microbiology, Marburg, Germany
[4]Center for Synthetic Microbiology, Marburg, Germany
[7]To whom correspondence should be addressed. Email: c.broedersz lmu.de

# Contents

## S1 Experimental determination of distances between loci 10 kb apart

### Experimental procedure

#### Bacterial strains and growth conditions

All *C. crescentus* strains used in this study were derived from the synchronizable wild-type CB15N (NA1000). Cells were grown in peptone-yeast extract (PYE) medium (Pointdexter, 1964) at $28°C$ under aerobic conditions (shaking at 210 rpm). When appropriate, the medium was supplemented with antibiotics at the following concentrations ($\mu$g/ml in liquid/solid medium): kanamycin (30/50), gentamycin (15/20), and spectinomycin (50/100).

#### Plasmid and strain construction

The bacterial strains, plasmids, and oligonucleotides used in this study are listed in Tables S1-S4. *Escherichia coli* TOP10 (Invitrogen) was used as host for cloning purposes. All plasmids were verified by DNA sequencing. Plasmids carrying 10 copies of either the lacO (stemming from the plasmid pLAU43 [5]) or the tetO array (stemming from the plasmid pLAU44 [5]) were integrated via single homologous recombination into *C. crescentus* via electroporation [2]. The genes for lacI-eCFP and tetI-eYFP (stemming from plasmid pLAU53 [5]) were integrated (single homologous recombination) at the xylX locus [12] via phage transduction mediated by phage $\phi$Cr30 [2]. Proper chromosomal integration was verified by colony PCR.

#### Microscopy and image analysis

All microscopy analyses were performed on cells grown in PYE Kn Gn till mid-exponential phase (OD 0.4), that were subsequently synchronized [9]. Immediately after synchronization, cells were immobilized on pads made of 1% agarose in PYE medium. Cells were observed with a Zeiss Axio Observer.Z1 microscope equipped with an alpha Plan-Apochromat 100x/1.46 Oil Ph3 objective (Zeiss, Germany). An X-Cite 120PC metal halide light source (EXFO, Canada), combined with ET-CFP and ET-YFP filter cubes (Chroma, USA), was used for detection of fluorescent foci. Pictures were taken with a pco.edge sCMOS camera (pco, Germany) and recorded with VisiView 2.1.4 (Visitron, Germany). To identify the subpixel localization of the fluorescent foci, a 2D Gaussian was fitted to each fluorescent focus using the GDSC SMLM plugin [11] [1] for ImageJ2 [7]. In order to correct for systematic shifts between the YFP and CFP channels, fiducials (Tetraspeck

---

[1] http://www.sussex.ac.uk/gdsc/intranet/microscopy/UserSupport/AnalysisProtocol/imagej/smlm_plugins/

2

Table S1: Strains used in this study.

| Strain | Genotype/description | Construction/Reference |
|---|---|---|
| **_E. coli_ strains** | | |
| TOP10 | Cloning strain | Invitrogen |
| **_C. crescentus_ strains** | | |
| CB15N | Synchronizable wild-type strain | Evinger & Agabian (1977) [3] |
| MvT151 | CB15N $P_{xyl}$::$P_{xyl}$- _lacI-ecfp-tetR-eyfp_ 10x _tetO_ and 10x _lacO_ spaced 10.0 kb apart at 196° | Subsequent integration of pMvT149, pMvT150 and $P_{xyl}$-_lacI-ecfp-tetR-eyfp_ into CB15N |
| MvT152 | CB15N $P_{xyl}$::$P_{xyl}$- _lacI-ecfp-tetR-eyfp_ 10x _tetO_ and 10x _lacO_ spaced 10.1 kb apart at 212° | Subsequent integration of pMvT151, pMvT152 and $P_{xyl}$-_lacI-ecfp-tetR-eyfp_ into CB15N |
| MvT170 | CB15N $P_{xyl}$::$P_{xyl}$- _lacI-ecfp-tetR-eyfp_ 10x _tetO_ and 10x _lacO_ spaced 10.1 kb apart at 21° | Subsequent integration of pMvT161, pMvT162 and $P_{xyl}$-_lacI-ecfp-tetR-eyfp_ into CB15N |
| MvT171 | CB15N $P_{xyl}$::$P_{xyl}$- _lacI-ecfp-tetR-eyfp_ 10x _tetO_ and 10x _lacO_ spaced 10.0 kb apart at 108° | Subsequent integration of pMvT163, pMvT164 and $P_{xyl}$-_lacI-ecfp-tetR-eyfp_ into CB15N |
| MvT172 | CB15N $P_{xyl}$::$P_{xyl}$- _lacI-ecfp-tetR-eyfp_ 10x _lacO_ and 10x _tetO_ spaced 10.0 kb apart at 108° | Subsequent integration of pMvT165, pMvT166 and $P_{xyl}$-_lacI-ecfp-tetR-eyfp_ into CB15N |
| MvT179 | CB15N $P_{xyl}$::$P_{xyl}$- _lacI-ecfp-tetR-eyfp_ 10x _tetO_ and 10x _lacO_ spaced 10.1 kb apart at 311° | Subsequent integration of pMvT159, pMvT160 and $P_{xyl}$-_lacI-ecfp-tetR-eyfp_ into CB15N |

microspheres, 0.5 $\mu$m, Invitrogen/Thermo Fischer Scientific, USA) were imaged in YFP and CFP channels and analyzed with the same set-up and pipeline.

Table S2: Plasmids used in this study.

| Plasmid | Description | Reference |
| --- | --- | --- |
| **Basic vectors** | | |
| pLAU43 | Plasmid carrying 240 copies of a *lacO* array, Kan$^R$ | Lau et al., 2003 [5] |
| pLAU44 | Plasmid carrying 240 copies of a *tetO* array, Gen$^R$ | Lau et al., 2003 [5] |
| pLAU53 | Plasmid carrying P$_{BAD}$-*lacI-ecfp tetR-eyfp*, Amp$^R$ | Lau et al., 2003 [5] |
| pMCS-2 | Integrating plasmid containing multiple cloning site, Kan$^R$ | Thanbichler et al., 2007 [8] |
| pMCS-4 | Integrating plasmid containing multiple cloning site, Gen$^R$ | Thanbichler et al., 2007 [8] |
| | | |
| **Plasmids constructed in this work** | | |
| pMvT149 | pMCS-2 including 10x tetO arrays and part of CCNA_02049, Kan$^R$ | This study |
| pMvT150 | pMCS-4 including 10x lacO arrays and part of gDNA close to CCNA_02054, Gen$^R$ | This study |
| pMvT151 | pMCS-2 including 10x tetO arrays and part of gDNA close to CCNA_02228, Kan$^R$ | This study |
| pMvT152 | pMCS-4 including 10x lacO arrays and part of gDNA close to CCNA_02233, Gen$^R$ | This study |
| pMvT159 | pMCS-2 including 10x tetO arrays and part of CCNA_03310, Kan$^R$ | This study |
| pMvT160 | pMCS-4 including 10x lacO arrays and part of gDNA close to CCNA_03317, Gen$^R$ | This study |
| pMvT161 | pMCS-2 including 10x tetO arrays and part of CCNA_00217, Kan$^R$ | This study |
| pMvT162 | pMCS-4 including 10x lacO arrays and part of gDNA close to CCNA_00226, Gen$^R$ | This study |
| pMvT163 | pMCS-2 including 10x tetO arrays and part of CCNA_01105, Kan$^R$ | This study |
| pMvT164 | pMCS-4 including 10x lacO arrays and part of gDNA close to CCNA_01112, Gen$^R$ | This study |
| pMvT165 | pMCS-2 including 10x lacO arrays and part of CCNA_01105, Kan$^R$ | This study |
| pMvT166 | pMCS-4 including 10x tetO arrays and part of gDNA close to CCNA_01112, Gen$^R$ | This study |

Table S3: Construction of plasmids.

| Plasmid | Description |
| --- | --- |
| pMvT149 | a) amplification of 10 tetO arrays from pLAU44 using oMvT789 & oMvT790 (product 433 bp) and 800 bp from NA1000 gDNA using oMvT791 & oMvT792 (product 848 bp) |
|  | b) fusion of two inserts with pMCS-2/NdeI+NheI via Gibson Assembly |
| pMvT150 | a) amplification of 10 lacO arrays from pLAU43 using oMvT796 & oMvT797 (product 547 bp) and 800 bp from NA1000 gDNA using oMvT798 & oMvT799 (product 845 bp) |
|  | b) fusion of two inserts with pMCS-2/NdeI+NheI via Gibson Assembly |
| pMvT151 | a) amplification of 10 tetO arrays from pLAU44 using oMvT803 & oMvT804 (product 435 bp) and 800 bp from NA1000 gDNA using oMvT805 & oMvT806 (product 843 bp) |
|  | b) fusion of two inserts with pMCS-2/NdeI+NheI via Gibson Assembly |
| pMvT152 | a) amplification of 10 lacO arrays from pLAU43 using oMvT808 & oMvT809 (product 548 bp) and 800 bp from NA1000 gDNA using oMvT810 & oMvT811 (product 845 bp) |
|  | b) fusion of two inserts with pMCS-2/NdeI+NheI via Gibson Assembly |
| pMvT159 | a) amplification of 10 tetO arrays from pLAU44 using oMvT789 & oMvT839 (product 435 bp) and 800 bp from NA1000 gDNA using oMvT840 & oMvT841 (product 843 bp) |
|  | b) fusion of two inserts with pMCS-2/NdeI+NheI via Gibson Assembly |
| pMvT160 | a) amplification of 10 lacO arrays from pLAU43 using oMvT819 & oMvT842 (product 549 bp) and 800 bp from NA1000 gDNA using oMvT843 & oMvT844 (product 851 bp) |
|  | b) fusion of two inserts with pMCS-2/NdeI+NheI via Gibson Assembly |
| pMvT161 | a) amplification of 10 tetO arrays from pLAU44 using oMvT789 & oMvT849 (product 436 bp) and 800 bp from NA1000 gDNA using oMvT850 & oMvT851 (product 840 bp) |
|  | b) fusion of two inserts with pMCS-2/NdeI+NheI via Gibson Assembly |
| pMvT162 | a) amplification of 10 lacO arrays from pLAU43 using oMvT819 & oMvT854 (product 549 bp) and 800 bp from NA1000 gDNA using oMvT855 & oMvT856 (product 844 bp) |
|  | b) fusion of two inserts with pMCS-2/NdeI+NheI via Gibson Assembly |
| pMvT163 | a) amplifation of 10 tetO arrays from pLAU44 using oMvT789 & oMvT859 (product 436 bp) and 800 bp from NA1000 gDNA using oMvT860 & oMvT861 (product 848 bp) |
|  | b) fusion of two inserts with pMCS-2/NdeI+NheI via Gibson Assembly |
| pMvT164 | a) amplification of 10 lacO arrays from pLAU43 using oMvT819 & oMvT863 (product 547 bp) and 800 bp from NA1000 gDNA using oMvT864 & oMvT865 (product 851 bp) |
|  | b) fusion of two inserts with pMCS-2/NdeI+NheI via Gibson Assembly |
| pMvT165 | a) amplification of 10 lacO arrays from pLAU43 using oMvT819 & oMvT867 (product 548 bp) and 800 bp from NA1000 gDNA using oMvT868 & oMvT861 (product 845 bp) |
|  | b) fusion of two inserts with pMCS-2/NdeI+NheI via Gibson Assembly |
| pMvT166 | a) amplification of 10 tetO arrays from pLAU44 using oMvT789 & oMvT869 (product 435 bp) and 800 bp from NA1000 gDNA using oMvT870 & oMvT865 (product 848 bp) |
|  | b) fusion of two inserts with pMCS-2/NdeI+NheI via Gibson Assembly |

Table S4: Oligonucliotides used in this study.

| ID | Name | Sequence (5' to 3') |
| --- | --- | --- |
| oMvT789 | tetO _CCNA_02049_p1 | cgagacgtccaattgcatatgtccctatcagtgatagagaggggaaagg |
| oMvT790 | tetO _CCNA_02049_p2 | cgccgctggccaccggatctctatcactgatagggaccttcccttctg |
| oMvT791 | tetO _CCNA_02049_p3 | gggaaggtccctatcagtgatagagatccggtggccagcggcgaac |
| oMvT792 | tetO _CCNA_02049_p4 | gatccccgggctgcagctagcgcgcactgaggccgatggcg |
| oMvT796 | lacO _CCNA_02049_p1 | gcgagacgtccaattgcatatgttgtgagcggataacaattggagcaag |
| oMvT797 | lacO _CCNA_02049_p2 | cttcgaccgctgggacttcttgttatccgctcacaatttgccttttgc |
| oMvT798 | lacO _CCNA_02049_p3 | ggcaaattgtgagcggataacaagaagtcccagcggtcgaagaggacg |
| oMvT799 | lacO _CCNA_02049_p4 | gatccccgggctgcagctagcgcctatgacgtgatgagctccaagcac |
| oMvT803 | tetO _CCNA_02228_p1 | cgagacgtccaattgcatatgtccctatcagtgatagagaggggaaagg |
| oMvT804 | tetO _CCNA_02228_p2 | gacgaccccctactggtcctctctatcactgatagggaccttccc |
| oMvT805 | tetO _CCNA_02228_p3 | ggtccctatcagtgatagagaggaccagtaggggggtcgtcgaacg |
| oMvT806 | tetO _CCNA_02228_p4 | gatccccgggctgcagctagcccagccccgccgccgacatcg |
| oMvT808 | lacO _CCNA_02228_p1 | gcgagacgtccaattgcatatgttgtgagcggataacaattggagcaag |
| oMvT809 | lacO _CCNA_02228_p2 | cccaggcaacttgtctttcgttgttatccgctcacaatttgccttttgc |
| oMvT810 | lacO _CCNA_02228_p3 | ggcaaattgtgagcggataacaacgaaagacaagttgcctgggc |
| oMvT811 | lacO _CCNA_02228_p4 | gatccccgggctgcagctagcctagcggatcgggcgcgcgaag |
| oMvT819 | lacO _CCNA_01737_p1 | gcgagacgtccaattgcatatgttgtgagcggataacaattggagcaag |
| oMvT839 | tetO _nusG_p2 | ggtcgaaaagatcgcctgatctctatcactgatagggaccttcccttc |
| oMvT840 | tetO _nusG_p3 | ggtccctatcagtgatagagatcaggcgatcttttcgacctgattg |
| oMvT841 | tetO _nusG_p4 | gatccccgggctgcagctagccgcgacagccgccgccgctcc |
| oMvT842 | lacO _CC-3211_p2 | gcagccgcgatttccattgagttgttatccgctcacaatttgccttttg |
| oMvT843 | lacO _CC-3211_p3 | ggcaaattgtgagcggataacaactcaatggaaatcgcggctgcgg |
| oMvT844 | lacO _CC-3211_p4 | ctagtggatccccgggctgcagctagcctgccaggagacgcggcc |
| oMvT849 | tetO _CC-0217_p2 | cagcgcatagcccagcgcgctctctatcactgataggacttccccttc |
| oMvT850 | tetO _CC-0217_p3 | ggtccctatcagtgatagagagcgcgctgggctatgcgctgac |
| oMvT851 | tetO _CC-0217_p4 | cccccgggctgcagctagcctagctccccgccctctcgatcg |
| oMvT854 | lacO _CC-0226_p2 | caactatgtcgatgacgagcattgttatccgctcacaatttgccttttg |
| oMvT855 | lacO _CC-0226_p3 | caaattgtgagcggataacaatgctcgtcatcgacatagttgctgcg |
| oMvT856 | lacO _CC-0226_p4 | ggatccccgggctgcagctagcgtgatgaccaagaccatgcttctggc |
| oMvT859 | tetO _CC-1053_p2 | gcccagatgccggcgcaatctctctatcactgatagggaccttcccttc |
| oMvT860 | tetO _CC-1053_p3 | gggaaggtccctatcagtgatagagagattgcgccggcatctgggcc |
| oMvT861 | tetO _CC-1053_p4 | gatccccgggctgcagctagcggcaggatcgaccaccgcgc |
| oMvT863 | lacO _CC-1059_p2 | ccagttcgcagagccggcgttgttatccgctcacaatttgccttttgc |
| oMvT864 | lacO _CC-1059_p3 | caaaaggcaaattgtgagcggataacaacgccggctctgcgaactggag |
| oMvT865 | lacO _CC-1059_p4 | ggatccccgggctgcagctagctcatgccatccggtagtgtcgggc |
| oMvT867 | lacO _CC-1053_p2 | gcccagatgccggcgcaatcttgttatccgctcacaatttgccttttgc |
| oMvT868 | lacO _CC-1053_p3 | ggcaaattgtgagcggataacaagattgcgccggcatctgggc |
| oMvT869 | tetO _CC-1059_p2 | ccagttcgcagagccggcgtctctatcactgatagggaccttcccttc |
| oMvT870 | tetO _CC-1059_p3 | ggaaggtccctatcagtgatagagacgccggctctgcgaactggag |

# S2 Data analysis: using experimental distance distributions to set the coarse-grained representation of the lattice polymer

We require a coarse-grained representation of the bacterial chromosome that is consistent with experimentally determined statistics beyond the coarse-graining length scale. Furthermore, our coarse-grained representation should allow for efficient computation. The resolution of the Hi-C data set (10 kb) sets a natural coarse-graining scale for the polymer, but we require additional experiments for the statistics at this length-scale: the distribution of spatial distances between pairs of loci at a 10 kb genomic distance. Here we demonstrate that a lattice polymer representation of the chromosome captures the statistics at this length scale. In this representation, the measured average spatial distance between a pair of loci sets the lattice spacing of our representation of the bacterial chromosome.

## S2.1 Analysis of experimental distance distributions of pairs of loci in *C. crescentus*

From the experimental procedure described in section S1, a data set of 100 2D distance vectors are obtained in *C. crescentus* for five pairs of loci separated by 10 kb. Note, microscopy data only gives us the projected 2D distances, while the actual distance vectors are in 3D. From the 2D data set, however, we can infer the underlying distribution of 3D distances. To make this inference, two effects are considered:

1. Measurement errors. This has two sources: finite localization precision and drift between the two consecutive measured images.

The measurement noise due to finite localization precision depends on the intensity of the fluorescent probe and the brightness of its direct surroundings. We calculated this precision using the GDSC SMLM plugin to have a standard error of 32.63 nm, with an average variation between measurements of 0.02 nm.

To account for drift between two consecutive images, we decompose the distance vector within each pair of foci into an $x$ and $y$ component, and sum the components for all cells. As the orientations of cells are isotropically distributed, these sums should go to 0 for increasing sample size. However, we find significant deviations from 0, larger than expected with our finite sampling, indicating a systematic drift. We correct for these deviations by subtracting the systematic drift in the $x$ and $y$ directions from each of the distance vectors. From the resulting distance vectors, a model for the 3D distance distribution is inferred. This correction will, however, be an overestimate: for a finite sample size, there will likely be deviations from 0, even in the absence of drift. To correct for this bias, we perform an iterative procedure: First, we simulate a measurement of a number of data points, drawn from the modeled 3D distribution without drift, equal to the sample size of the experiment. The simulated measurements of distance vectors are then shifted so that the mean $x$ and $y$ coordinates are exactly zero. This shift (which can be estimated accurately by simulating many realizations of the finite sample experiments) is added to all experimentally measured distances with our previous bias correction. We iterate this procedure until convergence is reached: the added shift in the last step must be equal to the expected shift of the inferred 3D distribution. Since this is only a small effect, convergence is reached for our data set after two iterations.

Table S5: Inferred average distances for the measured pairs of loci. The data sets MvT171 and MvT172 are for the same loci, just with their markers switched (see S1). The determined distances for each of these pairs are within two standard deviations of each other.

| Data set | Average 3D distance (nm) | Inferred $\sigma$ (nm) |
| --- | --- | --- |
| MvT151 | 106 ± 7 | 67 |
| MvT170 | 134 ± 8 | 84 |
| MvT171 | 121 ± 8 | 76 |
| MvT152 | 158 ± 9 | 99 |
| MvT172 | 132 ± 8 | 83 |
| MvT179 | 124 ± 7 | 78 |
| **Inferred average for entire chromosome** | Average 3D distance (nm) | Variance (nm) |
| | 129 ± 7 | 17 |

2. Intrinsic variations in three-dimensional distances between the loci, for instance due to thermal fluctuations of the DNA. We assume that the underlying distribution of relative positions is described by a 3D Gaussian with a standard deviation and a mean equal to 0. This results in one fit parameter ($\sigma$) for the underlying distribution.

To determine the value of $\sigma$ for each of the pairs of loci, we also use an iterative procedure: we start by choosing an initial value of $\sigma$, and then simulate the sampling of a large number of 3D distance vectors from this distribution. We then take a 2D projection of these samples and add the random measurement error of 32.63 nm (see point 1). Next, we compute the average 2D distance and compare with the experimentally determined 2D average distance. If these values are not equal, the value of $\sigma$ is updated accordingly, and a new round of the iteration begins. This procedure is repeated until convergence is reached (the average 2D distance is equal to the experimentally determined 2D average distance).

Once convergence is reached, the mean 3D distance for each pair of loci is calculated through a forward simulation of random points being drawn from a 3D Gaussian. The error on the mean inferred 3D distance for a specific pair of loci on the chromosome is determined by bootstrapping (see Table S5). The average distance for the entire chromosome is taken as the average over the means of the 5 pairs, and is determined to be $129 \pm 7$ nm (standard error of the mean).

Once the average distances are matched between model and experiment, the distributions of measured distances can also be compared. This distribution matches well between model and experiment (Fig. S1), supporting the assumption of a 3D gaussian as an underlying distribution of distances between the loci. Once we set the lattice constant of our lattice polymer to match this average 3D distance, our lattice polymer model approximately captures the correct Gaussian statistics for the distance between neighboring chromosomal regions. This validates the use of a lattice polymer to connect consecutive monomers representing neighboring chromosomal regions.

## S2.2 Setting the dimensions of the lattice spacing and the cellular confinement in the model

We employ a polymer model on a cubic lattice. In this representation, the position of each even monomer indicates the cubic unit cell occupied by the center of a Hi-C chromosomal region. Two monomers are defined to be in contact if they simultaneously occupy the same lattice site. This assumes that the dominant contributions to contacts between two chromosomal regions are from configurations where their respective centers occupy the same unit cell.

To set the scale of the lattice spacing $b$ in the model, we use the average spatial distance between consecutive Hi-C chromosomal regions determined in Sec. S2.1. If we consider two neighboring segments between chromosomal regions, however, coarse-graining effects need to be taken into account: only seven distances between these regions are possible in the lattice representation ($0$, $\sqrt{2}b$, $2b$, $\sqrt{6}b$, $\sqrt{8}b$, $\sqrt{10}b$, $4b$), which occur with respective relative occurrence frequencies ($f_1, \cdots, f_7$). In our MaxEnt model, we robustly observe ($f_1 \approx 0.092$, $f_2 \approx 0.50$, $f_3 \approx 0.13$, $f_4 \approx 0.19$, $f_5 \approx 0.041$, $f_6 \approx 0.048$, $f_7 \approx 0.0022$). This coarse-graining effect implies a cut-off of the tail of the underlying Gaussian distribution of 3D distances. To account for this cut-off, we first sample real-space configurations of consecutive chromosomal regions according to the experimentally determined 3D Gaussian distribution of continuous distances (see Sec. S2.1), and infer the statistics in the corresponding lattice model. For each of the seven possible (discretized) distances in the coarse-grained lattice representation, we thus obtain associated conditional distribution of real-space distances. The sum of the seven conditional real-space distance distributions, weighted by their respective relative occurrence frequencies ($f_i$), defines the full distribution of distances between neighbouring chromosomal regions in the MaxEnt model. We determine the lattice spacing $b = 88$ nm, such that the average distance between chromosomal regions in our MaxEnt model matches the experimentally determined average distance (Sec. S2.1)). Note, for this lattice spacing, the distribution of distances between neighbouring chromosomal regions in the MaxEnt model are also in accord with our experimentally determined distributions (Fig. S1).

To restrict the phase space of chromosome configurations to those that fit inside a cell, we introduce a confinement formed by a cylinder capped by two hemispheres. The dimensions of the confinement are chosen to match typical dimensions of a newborn swarmer cell. These dimensions are determined by taking a sample of 267 cells from the MvT151 data set, which yields an average length of $2.3 \pm 0.2, \mu m$ and width of $0.75 \pm 0.04 \mu m$. Subtracting the estimated width of the cell envelope of 61 nm (based on figure 2 of [4]), we arrive at typical chromosome confinement dimensions of $2.2 \times 0.63 \mu m$. With the inferred lattice spacing, this translates to a confinement of 470 unit cells (25 lattice spacings long and 7 wide). This representation of the cell could be refined further to include the crescent shape, but we find that such corrections do not appear to significantly affect the results of our model (see S12).

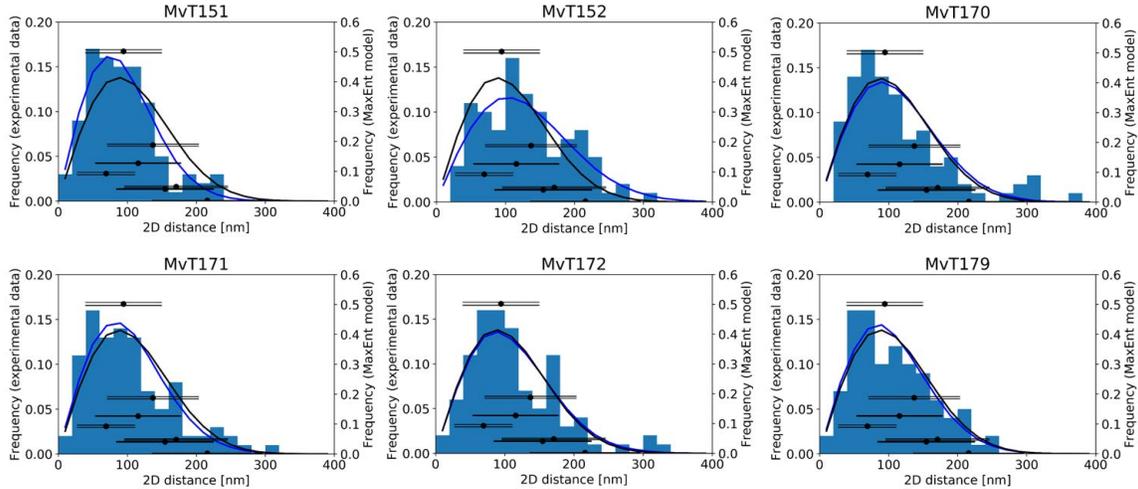

Figure S1: **Distributions of 2D projected distances from experiment and MaxEnt model.** Bars: experimentally measured 2D distances (after bias correction, see Sec. S2.1). Blue lines: distributions of 2D projected distances from the inferred 3D Gaussian distribution. For each data set there is one fit parameter $\sigma$, chosen such that the average distances of measured and inferred distributions match. Black markers: Relative frequencies ($f_i$) of each of the seven possible configurations of two neighboring chromosomal regions of the MaxEnt model with associated average distances determined from coarse-graining. The pairs of horizontal black lines at each dot indicate the mean variance of the MaxEnt configuration frequency for all neighboring pairs of chromosomal regions. The error bar indicates the standard deviation of the underlying distance distribution for each coarse-grained configuration. Black curve: Inferred 2D distance distribution between consecutive genomic regions for the entire chromosome for the MaxEnt model. This distribution is obtained by weighing the inferred distance distribution for each coarse-grained configuration with the associated relative occupancy frequency within the MaxEnt model. To enable a direct comparison with experimental data, the inferred measurement noise is applied over the MaxEnt distance distribution. Note that all MaxEnt data sets are the same in each panel.

## S3 Inverse Monte Carlo algorithm for MaxEnt chromosome model

We solve the inverse problem and obtain the Lagrange multipliers $\epsilon_{ij}$'s by an iterative procedure: we perform a Monte Carlo (MC) simulation (forward algorithm) to sample equilibrium states from the lattice polymer model with an initial guess for $\epsilon_{ij}$. Subsequently, we compare the estimated contact map, $f_{ij}^{\text{sim}}$, obtained from this MC simulation, with the target experimental map $f_{ij}^{\text{expt}}$. When the modeled and experimental contacts deviate, the $\epsilon_{ij}$'s are updated (inverse algorithm). This procedure converges when the modelled normalized contact frequency map matches the Hi-C data set within a tolerance level, yielding the complete set of parameters $\epsilon_{ij}$ that defines the MaxEnt model. The forward and inverse algorithm are described below.

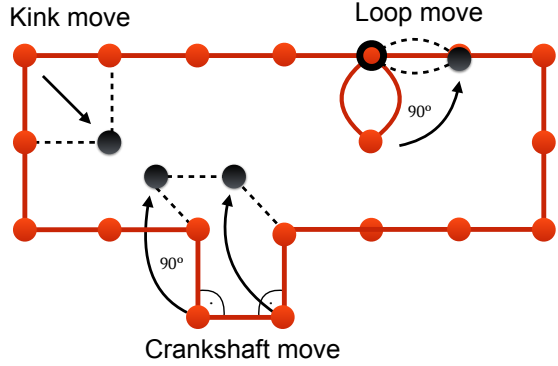

Figure S2: Illustration of the three polymer moves employed in the Monte Carlo simulation.

## S3.1 Forward algorithm

In our coarse-grained model, the bacterial chromosome of *C. crescentus* is represented by a circular lattice polymer with a length of 1620 monomers. Each $4^{\text{th}}$ monomer represents the location of the center of a genomic region, with three monomers in between to ensure Gaussian statistics between subsequent centers of genomic regions (see Sec. S2). The level of coarse-graining can be adapted to accommodate the resolution of the data on which the model is strained.

The algorithm is initiated with the polymer randomly arranged within the confinement. We simulate the Boltzmann distribution of polymer configurations in the MaxEnt model using Monte Carlo simulations. To sample configurations in the Monte Carlo algorithm, we employ three different polymer moves; the *kink move*, the *Crankshaft move* and the *loop move* (Fig. S2). This move set allows an ergodic sampling of the space of polymer configurations, which is demonstrated in Sec. S3.3.

A potential move $\{\mathbf{r}\} \to \{\mathbf{r}'\}$ is randomly chosen (based on the move set in Fig. S2), and then accepted with a probability $P_{\text{acc}}(\{\mathbf{r}'\}, \{\mathbf{r}\})$ according to the Metropolis criterion: $P_{\text{acc}}(\{\mathbf{r}'\}, \{\mathbf{r}\}) = \min(1, \exp(E(\{\mathbf{r}\}) - E(\{\mathbf{r}'\})))$, provided the configuration stays within the confinement. Here, $E(\{\mathbf{r}'\})$ and $E(\{\mathbf{r}\})$ are the energies of the proposed configuration $\{\mathbf{r}'\}$ and current configuration $\{\mathbf{r}\}$, respectively. The energies are computed according to the Hamiltonian (Eq. (5) in main text)

$$H(\{\mathbf{r}\}) = \frac{1}{2} \sum_{ij} \epsilon_{ij} \delta_{\mathbf{r}_i, \mathbf{r}_j}. \tag{S1}$$

## S3.2 Inverse algorithm

As noted in the main text, we learn the MaxEnt model directly from the normalized experimental Hi-C map. During a forward simulation of the polymer, the contact frequency $f_{ij}^{\text{model}}$ of each pair of monomers is counted. After one round of forward simulation, the simulated contact frequencies are normalized and compared to the experimental ones. The pairwise interaction energies are then updated according to

$$\Delta \epsilon_{ij} = \alpha(\tilde{f}_{ij}^{\text{model}} - \tilde{f}_{ij}^{\text{exp}}) \times \frac{1}{\sqrt{\tilde{f}_{ij}^{\text{exp}}}}. \tag{S2}$$

Here, $\alpha$ is the learning rate (which we typically set to 0.2), and the last factor is included to speed up conversion for pairs with a low contact frequency. Note, $\tilde{f}_{ij}^{\text{model}}$ and $\tilde{f}_{ij}^{\text{exp}}$ are the normalized model and experimental contact frequencies, respectively.

Importantly, to impose that the normalized contact frequencies match between model and experiment, we need to determine one remaining parameter: the absolute scale of the model contact frequencies. To fix this scale, we ensure that condition Eq. 6 (main text) is satisfied in each iteration step. This is done by applying an overall shift in the interaction energies after the update step in Eq. (S2). This overall shift can be derived as follows: we start from Eq. 6, which imposes $\sum_{ij} \epsilon_{ij} \tilde{f}_{ij}^{\text{expt}} = 0$. In general, a set of $\epsilon_{ij}$ obtained after the update step in Eq. (S2) will not satisfy this constraint. We can, however, introduce a shift $\Delta\epsilon$ of all $\epsilon_{ij}$ such that this condition is satisfied:

$$\sum_{ij} (\epsilon'_{ij} - \Delta\epsilon) \tilde{f}_{ij}^{\text{expt}} = 0. \tag{S3}$$

Rewriting, and making use of $\sum_{ij} \tilde{f}_{ij}^{\text{expt}} = N_{\text{bin}}$ with $N_{\text{bin}}$ is the number of Hi-C bins, yields

$$\Delta\epsilon = -\frac{\sum_{kl} \epsilon'_{kl} \tilde{f}_{kl}^{\text{exp}}}{N_{\text{bin}}}. \tag{S4}$$

Performing this shift after each update step ensures that the condition in Eq. (8) is satisfied at each iteration of the inverse algorithm.

We iterate the inverse algorithm until the Pearson's correlation coefficient between the simulated normalized contact frequencies and the experimental data is above 0.98. This is the correlation coefficient of contact frequencies between repeat experiments reported in [6]. In practice, we can obtain even higher correlation coefficients of 0.9996, as stated in the main text. With each subsequent forward simulation, the number of Monte Carlo steps is multiplied by $\sqrt{n}$, with $n$ the iteration step. The inverse algorithm is typically started with $\sim 360$ million steps, and run for $\sim 100$ iterations.

### S3.3 Ergodicity of forward algorithm

Next, we demonstrate that the algorithm is ergodic. A circular path of the polymer can be represented as a sequence of $N$ steps along the lattice, where each step is either up ($U$), down ($\bar{U}$), right ($R$), left ($\bar{R}$), in ($I$) or out ($\bar{I}$). We denote the total number of steps of type $x$ by $N(x)$. Circularity of the path implies that $N(U) = N(\bar{U})$, $N(R) = N(\bar{R})$ and $N(I) = N(\bar{I})$. Furthermore, we will divide the steps in *types*, where ($U$) and ($\bar{U}$) are type 1, ($R$) and ($\bar{R}$) are type 2, and ($I$) and ($\bar{I}$) are type 3. An individual path can then be described as a sequence of steps, for example

$$[\text{U}, \bar{\text{R}}, \text{I}, \text{R}, \bar{\text{U}}, \cdots]. \tag{S5}$$

Here, each of the steps is colored by type. In the following we will also consider the sequence within each type. For our example, the sequences for the three types are:

- Type 1: $[\text{U}, \bar{\text{U}}, \cdots]$
- Type 2: $[\bar{\text{R}}, \text{R}, \cdots]$
- Type 3: $[\text{I}, \cdots]$

We now consider the action of each of the polymer moves on a sequence of steps.

- The **kink move** interchanges two subsequent steps of a different type. Using only this move, any sequence of type 1, type 2 and type 3 steps can be created from a starting sequence that doesn't change the number of each type. Put differently, using the representation in (S5), any sequence of red, green and blue can be created that conserves the original counts of each color. Within each type, the sequence of the possible steps (e.g. $U$ and $\bar{U}$), however, cannot be changed with this move.

- The **crankshaft move** takes a motif of the form $[A, B, \bar{A}]$ and alters this to one of three possible motifs: (i) $[\bar{A}, B, A]$, or (ii) $[C, B, \bar{C}]$, or (iii) $[\bar{C}, B, C]$. The first alteration changes the sequence of steps within a type. Combining this alteration with the kink move, any sequence of steps within each type can be created, provided that there is at least one set of steps of a different type.

  Alteration (ii) and (iii) change the number of steps of each type: $N(A) + N(\bar{A})$ is reduced by 2, and $N(C) + N(\bar{C})$ is increased by 2. Combining this with the kink move, any set of counts of each of the types can be created, provided that polymer length and circularity are preserved, and that in the initial state not all steps are of the same type.

  Combining all three alterations with the kink move, from any starting sequence any final sequence can be created that conserves polymer length and circularity, as long as the starting and final sequence have moves of at least two different types.

- The **loop move** takes a motif of the form $[A, \bar{A}]$ and alters it to either (i) $[\bar{A}, A]$ or (ii) $[B, \bar{B}]$ or (iii) $[\bar{B}, B]$. Alteration (i) enables any change of the sequence within a type when the entire initial sequence is of the same type. Alterations (ii) and (iii) allow the conversion from a state of only one type to a state of two types.

  Combining the loop move with the kink and crankshaft moves, from any starting sequence any final sequence can be created that conserves polymer length and circularity. Thus, an ergodic sampling of the space of polymer configurations is ensured.

Note I: The presence of a confinement introduces a parity on the lattice sites: sites that can be occupied by an even monomer through these 3 moves cannot be occupied by an uneven monomer, and vice versa. Either choice of parity can be seen as a separate coarse-grained model, as the unit cell locations shift depending on this choice.

Note II: A confinement could be chosen that 'traps' a portion of the polymer in place, making the phase space reachable using the three moves dependent on the initial state. For our confinement consisting of a cylinder with rounded edges such a trapping is not present, thus ergodicity is still preserved.

Note III: ergodicity is already ensured if only the loop and kink moves are used: the crankshaft move can be constructed as a combination of the two. However, the crankshaft move allows for a faster exploration of phase space and is thus also included.

## S4 Testing the inverse Monte Carlo algorithm

To test the performance of our inverse algorithm, we generated trial data sets by running a forward simulation for a chosen set of input effective interaction energies $\epsilon_{ij}^{\text{in}}$ (upper left Fig. S3A). The

resulting simulated contact map, $f_{ij}^{\text{in}}$, exhibits intricate features, including domain-like structures along the main diagonal and a fainter second diagonal (upper left Fig. S3B). Subsequently, we treat this contact map as an experimental data set, which we use as an input to our iterative inverse scheme. We find that our inverse scheme rapidly and accurately retrieves the correct energies, $\epsilon_{ij}^{\text{model}} \approx \epsilon_{ij}^{\text{in}}$, and contact frequencies, $f_{ij}^{\text{model}} \approx f_{ij}^{\text{in}}$, demonstrating that this scheme adequately solves the inverse problem (Fig. S3A-C).

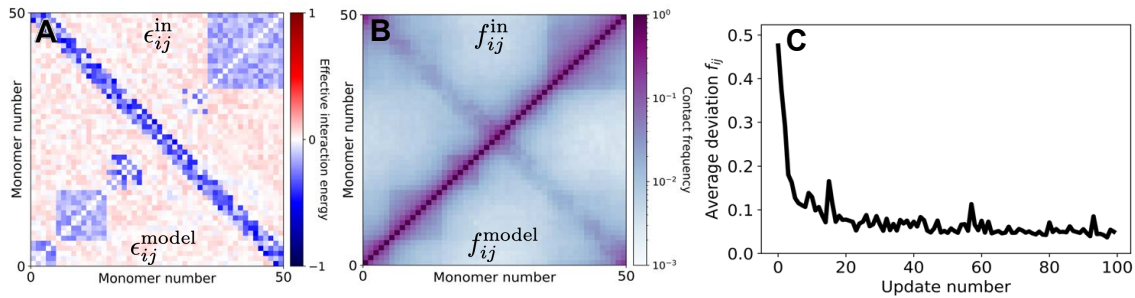

Figure S3: **Demonstration of numerical inverse algorithm for MaxEnt chromosome model.** **A** Upper left: input effective interaction energies $\epsilon_{ij}^{\text{in}}$. Lower right: effective interaction energies retrieved by the MaxEnt model. **B** Upper left: simulated contact frequencies $f_{ij}^{\text{in}}$ using $\epsilon_{ij}^{\text{in}}$. Lower right: contact frequencies of the MaxEnt model, using $f_{ij}^{\text{in}}$ as an input. **C** The average relative contact frequency deviation: $\langle f_{ij}^{\text{in}} - f_{ij}^{\text{model}} \rangle / \langle f_{ij}^{\text{model}} \rangle$ vs. iteration number of inverse algorithm.

## S5  Hi-C data processing

Before the Hi-C data from Ref. [6] can be used to train our MaxEnt model, we need to account for a known artefact: there is a spurious local increase in contact frequencies between the ori and ter genomic regions. Although this increase is not readily visible on a linearly scaled Hi-C map (Fig. 1), it can be easily seen on a logarithmic scale (Fig. S4A). This local increase has been assumed to be due to a small fraction of cells having started replication, and the newly formed ori regions having partially crossed over to the ter-side of the cell [10].

When inter-arm contact frequencies are plotted versus genomic distance on a log-log plot, this effect gives rise to a sudden, sharp increase around large genomic distance (Fig. S4A). To remove this artefact from the data, we separately consider two regimes: inter-arm distances below (regime I) and above (regime II) the minimum in the average contact frequencies (red line Fig. S4A)). From extrapolation of the data in regime I, we expect that the data in regime II is completely dominated by spurious ori-ter interactions since we would not be able to distinguish the real contacts in regime II from zero due to the noise floor. Therefore, we correct the data in regime II by setting the contacts equal to zero. We can use the measured data in regime II to also correct the data in regime I to avoid biases in our model due to any spurious *ori-ter* interactions. For each genomic distance bin in regime II, an average and standard deviation is calculated. We assume that the contact frequencies resulting from interactions between the old ter and the new *ori* regions increase with genomic distance according to a power law scaling. Thus, we can perform a linear fit over the log-log plot in regime II. Here, we fitted the curve to two standard deviations above the average

14

(roughly the upper envelope of the data points) to make sure we have a good estimate of the upper-bound on spurious counts. We then subtract this upper bound from all points in regime I. We then reapply the normalization procedure used in [6] to the modified data set.

Note, this correction only makes a significant difference for the data in a small part of region I near the transition to region II. For genomic regions on opposite chromosomal arms, the same procedure is applied. This correction results in a contact frequency map where the main diagonal and intra-arm contact frequencies are largely unaffected, but the local increase in contact frequencies between *ori* and *ter* regions is reduced (Fig. S5). We use this corrected contact frequency map to train the MaxEnt model.

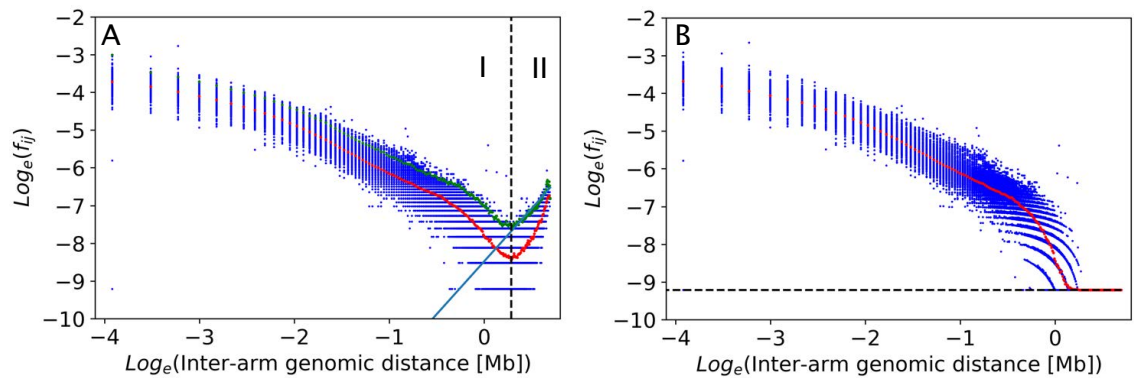

Figure S4: **Correction of contact frequencies between *ori* and *ter* regions.** A: Log-log plot of inter-arm Hi-C scores before correction (blue dots), together with the average distance per bin (red dots) and the average plus two standard deviations (green dots). The light blue line indicates a fit through the green dots after the inflection point (black dashed line). B: Hi-C scores after correction (blue dots) together with the average per genomic distance bin (red dots).

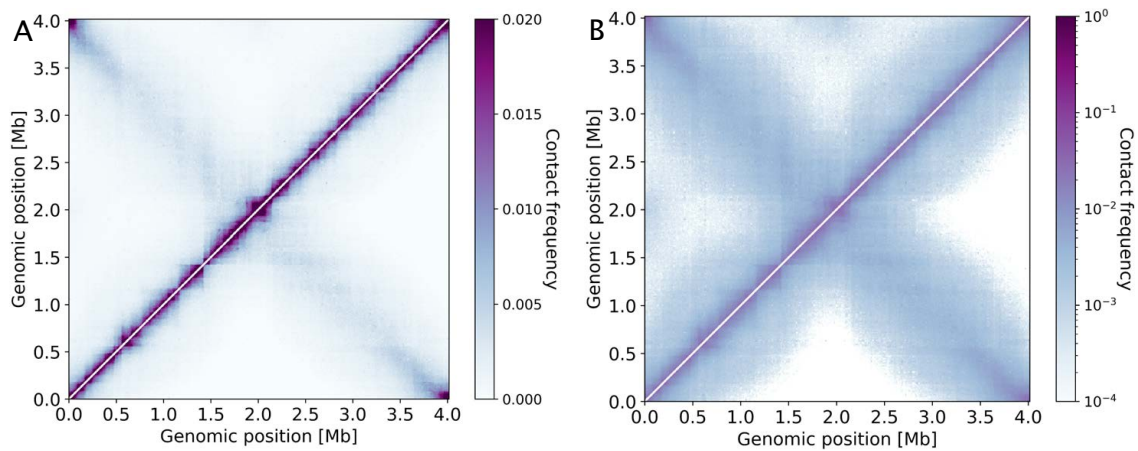

Figure S5: Hi-C scores before correction (upper left triangle), and after correction (lower right triangle) on a linear scale (A) and a logarithmic scale (B).

## S6 Analysis of genomic Super Domains

To define genomic Super Domains (SuDs), we first choose a cluster radius $r$. For each genomic region $i$, we consider a specific configuration of the chromosome and then calculate the length $\ell$ of the set of subsequent genomic regions (in both directions along the chromosome) that lie within the radius $r$ from the position of genomic region $i$ (illustrated by the black line in Fig. 3D). We observe that for each configuration of the chromosome, the genomic regions separate into a small number of domains, indicated by the blue and red areas in Fig. 3D. We identify a domain with each local maximum in $\ell$ (indicated by $L1 - L3$ and $R1 - R3$ in Fig. 3E); the peak location represents the genomic region at the center of a SuD and the peak value indicates the number of genomic regions within the domain.

To determine a natural choice for $r$, we perform a parameter sweep over $r$ and consider the change in the average value of $\ell$ with $r$: $d\bar{\ell}/dr$. We find that for the MaxEnt models on wild-type, rifampicin treated and $\Delta smc$ cells, $d\bar{\ell}/dr$ initially increases with $r$, and then becomes approximately constant (Fig. S6). For models unconstrained by Hi-C data (the 'random polymer', and the 'tethered random polymer'), such a transition to a plateau regime is not present. We interpret the transition to this plateau regime in the MaxEnt models as the genomic length scale at which the linear organization of the chromosome along the cell length starts dominating local fluctuations of loci(See Fig. 2A&B). We take the crossover point between these two regimes to be $r = 290$ nm, indicated by the grey dashed line in Fig. S6B.

To quantify the degree of long-axis exclusion between SuDs, the distribution of long-axis positions of the genomic regions contained in each domain is computed (Fig. 3E). A long-axis position is assigned to a Super Domain based on the highest-occupied long axis coordinate of this cluster. The degree of overlap of long-axis positions is then computed for randomly paired left and right arm configurations and for correctly matched pairs.

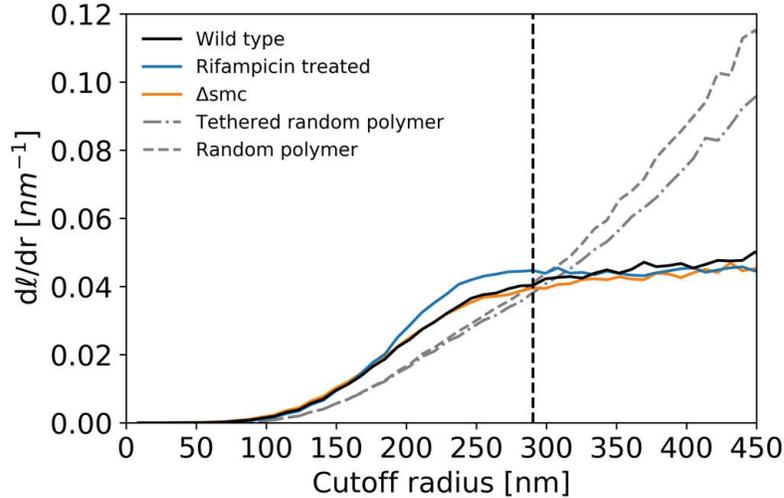

Figure S6: **Cluster analysis.** Derivative of the average cluster size as a function of the cutoff radius $r$, for wild-type cells (black), rifampicin treated cells (blue), a $\Delta smc$ mutant (orange), a tethered random polymer (dash-dotted line) and a random polymer (dashed line). The vertical dashed line indicates the chosen cutoff value.

## S7 Overlap analysis between local chromosome extension peaks and highly transcribed genes

To investigate the connection between peaks in the local extension profile and the locations of highly transcribed genes, we first construct a (nonlinear) trend line through the chromosome extension profile. This line is constructed by repeatedly applying a Gaussian smoothing filter over the data, incorporating periodic boundary conditions. The Gaussian smoothing is implemented by repeatedly applying a moving average over groups of 3 subsequent genomic regions. We find that 250 repeats to result in a satisfactory balance between smoothing out local peaks and keeping the larger-scale trend (grey line in Fig. S7A). Next, we select the subset of local extension peaks that lie a factor $\alpha$ above the trend line. We perform a sweep over $\alpha$ and calculate for each choice of $\alpha$ the fraction of incorporated peaks that coincide with the locations of highly transcribed genes. Additionally, for each $\alpha$ we simulate a number of randomly positioned peaks equal to the number of incorporated peaks. From this simulation, we calculate the expected fraction of overlap and the 95% confidence intervals.

We find that the fraction of overlap is significantly higher than expected for randomly positioned local extention peaks, if up to the 9 highest peaks are considered (Fig. S7B). If more peaks are incorporated, the fraction of overlap gradually decays to the level expected for random positions. Repeating this analysis for the right (0-2 Mb) and left (2-4 Mb) chromosomal arms seperately, we find that the fraction of overlap is only significantly higher than a random guess for the highest peaks of the right arm (Fig. S7C). For the left arm, by contrast, the fraction of overlap is close to the value expected by random guess for all values of $\alpha$.

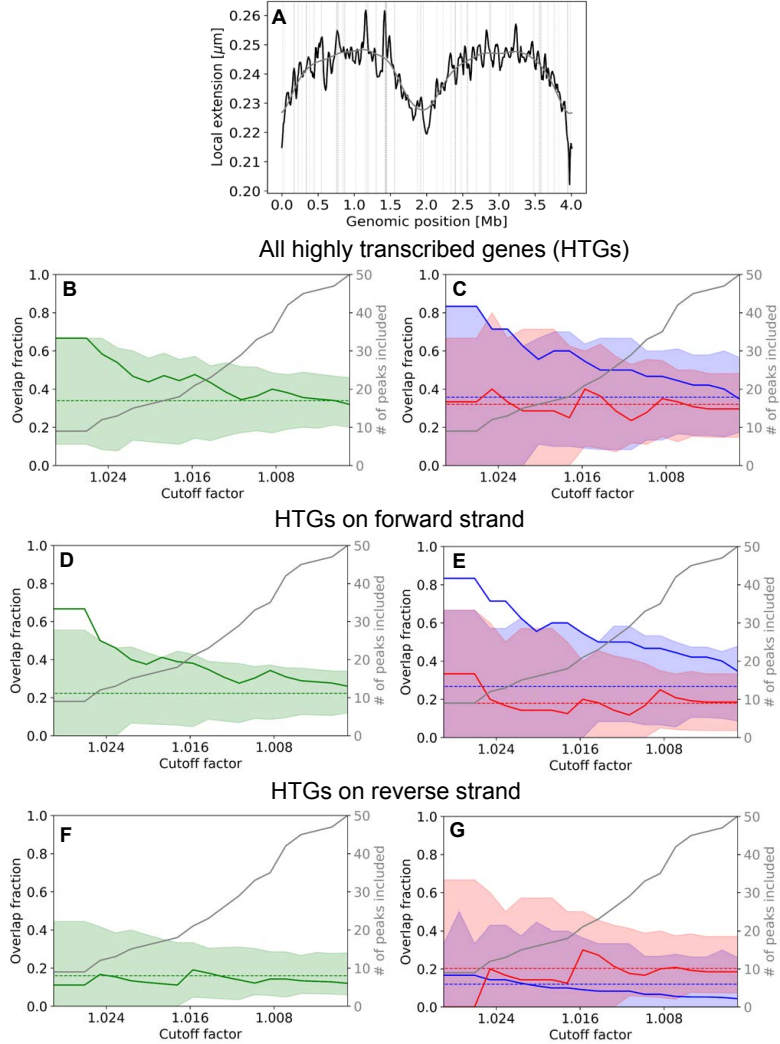

Figure S7: **Analysis of the degree of overlap between peaks in local chromosome extension and the locations of highly transcribed genes.** **A** Wild-type local chromosome extension profile (black line), together with a trend line obtained from Gaussian smoothing (grey line) and the locations of highly transcribed genes (HTGs) (vertical dashed lines). **B** Green solid line: fraction of local extension peaks that coincide with the location of a highly transcribed gene, as a function of the cutoff factor $\alpha$. The dashed line indicates the expected fraction of overlap for randomly chosen locations of peaks, the light green area indicates the 95% confidence interval around this expected fraction. The grey line indicates the number of peaks included for a given cutoff factor (indicated on the right axis). **C** The same analysis as in B, performed separately for the right (0-2 Mb, blue) and left (2-4Mb, red) chromosomal arms. **D,E** The same analyses as in B and C, using only the positions of HTGs located on the forward strand of the chromosome. **F,G** The same analyses as in B and C, using only the positions of HTGs located on the reverse strand of the chromosome.

18

# S8 Relation between Hi-C scores and average distance and distance correlations

Previous modelling approaches for the *C. crescentus* chromosome used average distance based models to find typical chromosome configurations [10, 13]. In these approaches, an experimentally determined average linear relation between intra-arm genomic distances and average spatial distances was used to derive a functional relation between Hi-C contact scores and average spatial distances. Our MaxEnt model does not require this assumption, instead we can use the model to predict the relation between Hi-C scores and average distances. Interestingly, our MaxEnt model predicts an approximately linear relation between Hi-C scores and average distances, but with significant deviations from this average trend for individual pairs of genomic regions (Fig. S9A). Moreover, there are substantial deviations from a linear trend for small and large genomic distances. Finally, we also observe significant variations around an average trend for Hi-C scores versus spatial distances (Fig. S9B).

In addition to these variations in average spatial distances, we also find significant correlations in deviations from these averages for individual configurations throughout the entire chromosome (Fig. S9). In previously used approaches [10, 13] such correlations could not be taken into account, which could explain the difference in predictions from our MaxEnt model.

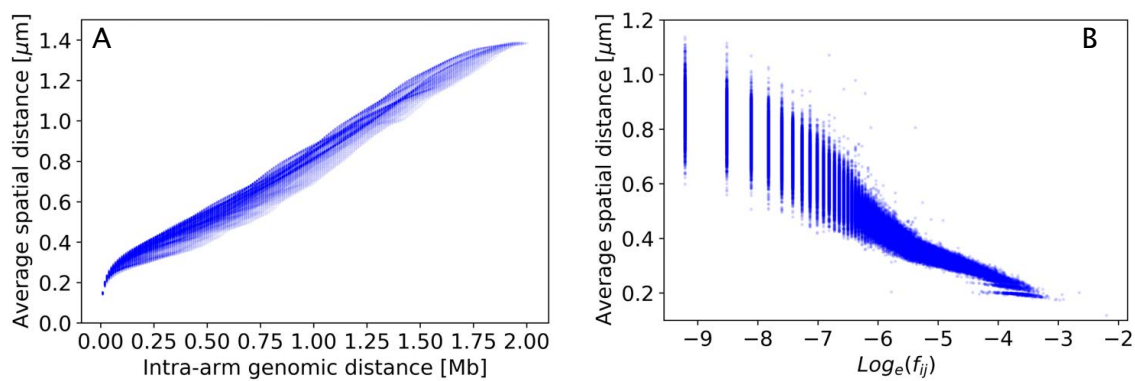

Figure S8: **Variations of average distance statistics between individual pairs of genomic regions.** **A** Average spatial distance versus genomic distance predicted by the MaxEnt model. **B** Average spatial distance versus the logarithm of the Hi-C score predicted by the MaxEnt model.

19

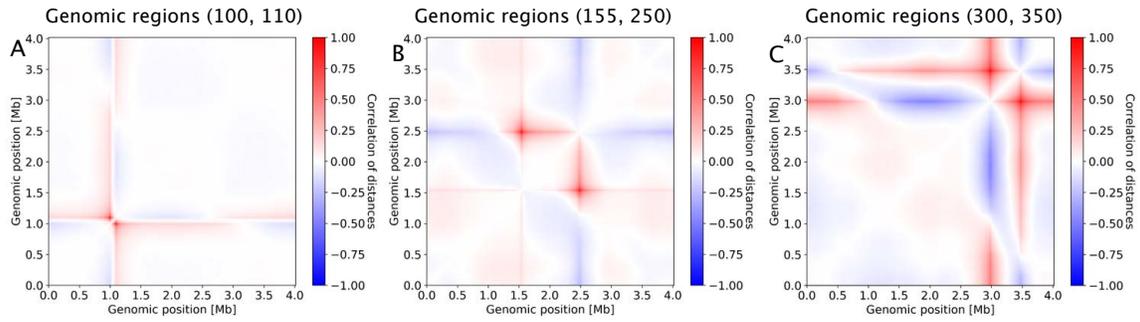

Figure S9: Correlations between distances of all pairs of genomic regions, and the distance between three sample pairs.

## S9    MaxEnt models for $\Delta smc$ cells and rifampicin treated cells

We apply the same approach to perform a Hi-C data analysis and MaxEnt model inference for rifampicin treated cells and $\Delta smc$ cells. The prepossessing of Hi-C data is shown in Figs. SS10 and SS11, and the corresponding MaxEnt models are shown in Figs. SS12 and SS13. We show the results for the long-axis localization in Fig. S14 together with previously published experimental data, and various correlation functions are depicted in Fig. S15.

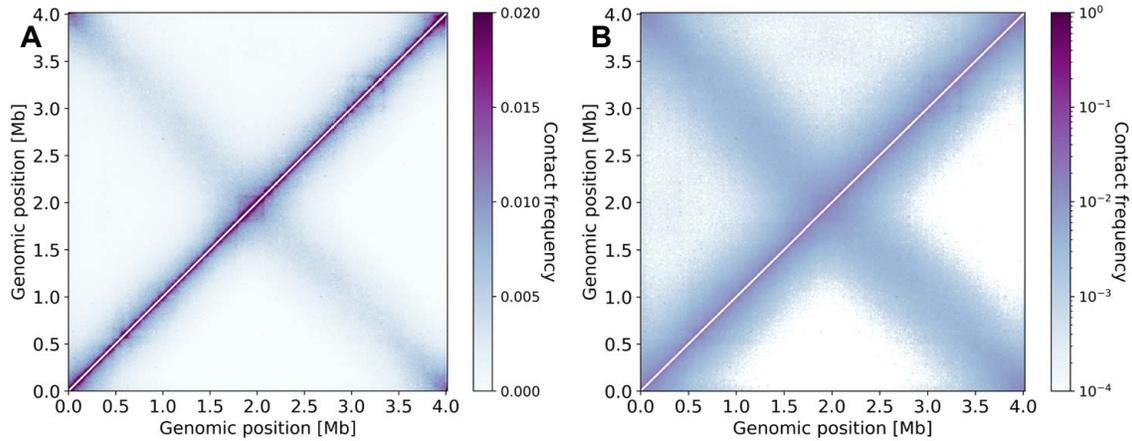

Figure S10: Hi-C scores of rifampicin treated cells before correction (upper left triangle), and after correction (lower right triangle) on a linear scale (A) and a logarithmic scale (B).

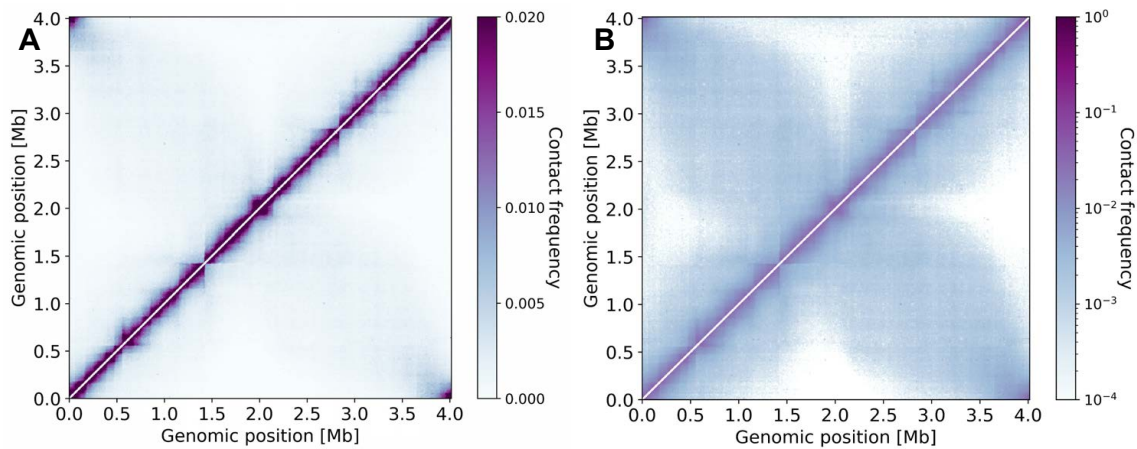

Figure S11: Hi-C scores of $\Delta smc$ cells before correction (upper left triangle), and after correction (lower right triangle) on a linear scale (A) and a logarithmic scale (B).

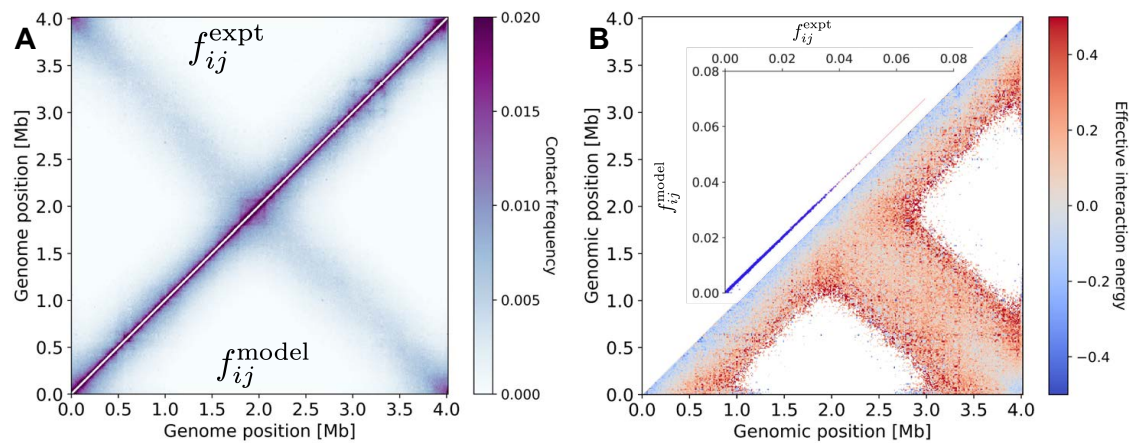

Figure S12: Rifampicin treated cells: comparison between $\Delta smc$ experimental contact frequencies $f_{ij}^{\mathrm{expt}}$ (upper left corner, adapted from Ref. [6]) and contact frequencies obtained from our inferred MaxEnt model $f_{ij}^{\mathrm{model}}$ (lower right corner). **B** Inferred effective interaction energies $\epsilon_{ij}$ (lower right corner) together with scatterplot of $f_{ij}^{\mathrm{expt}}$ vs. $f_{ij}^{\mathrm{model}}$ (inset).

21

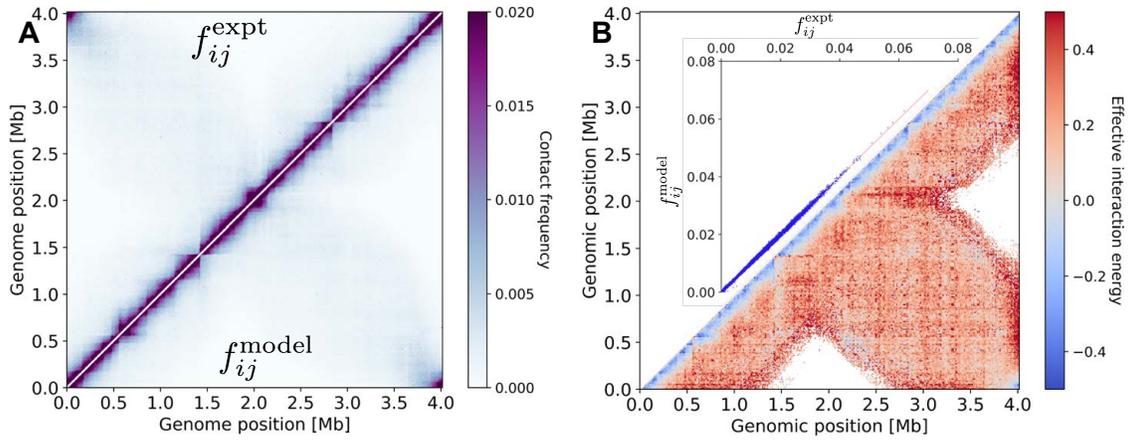

Figure S13: $\Delta smc$ cells: comparison between experimental contact frequencies $f_{ij}^{\text{expt}}$ (upper left corner, adapted from Ref. [6]) and contact frequencies obtained from our inferred MaxEnt model $f_{ij}^{\text{model}}$ (lower right corner). **B** Inferred effective interaction energies $\epsilon_{ij}$ (lower right corner) together with scatterplot of $f_{ij}^{\text{expt}}$ vs. $f_{ij}^{\text{model}}$ (inset).

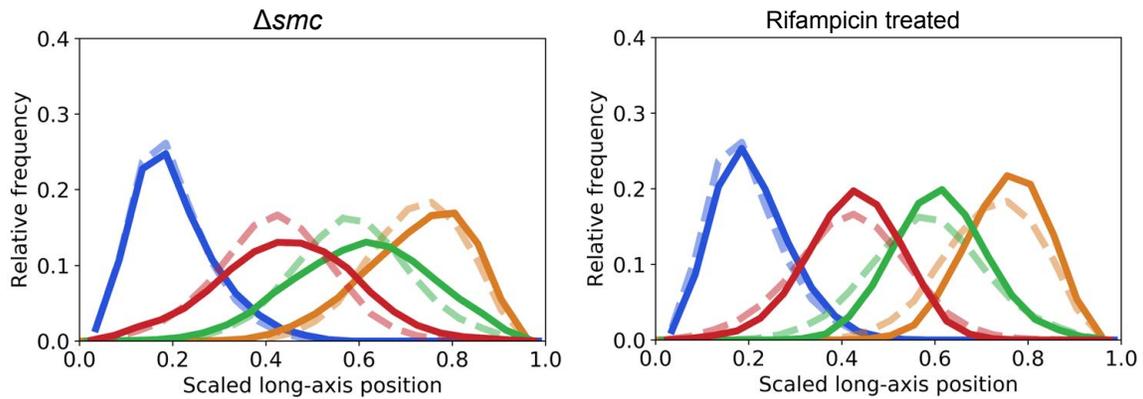

Figure S14: Comparison between inferred long-axes localization distributions for wild-type cells (dashed lines) and $\Delta smc$ mutants (solid lines left) and rifampicin treated cells (solid lines right).

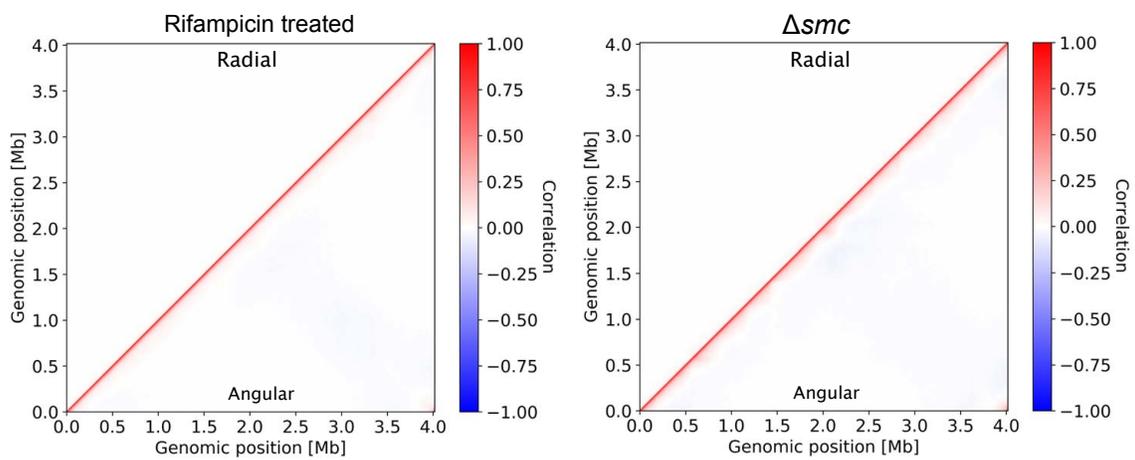

Figure S15: Correlations in the radial positions (upper left corner) and orientations around the long axis (lower right corner) between all pairs of genomic regions, for rifampicin treated cells (left) and $\Delta smc$ mutants (right).

## S10 Estimates of localization information

To compute the localization information for a genomic region, we first calculate the average occupation $P^{s,i}$ of each unit cell $s$ for each genomic region $i$ during a forward simulation. The localization entropy $S^i_{\text{loc}}$ in bits of site $i$ is then calculated by [1]

$$S^i_{\text{loc}} = \sum_s P^{s,i} \log_2 P^{s,i}. \tag{S6}$$

The positional information is calculated by subtracting $S^i_{\text{loc}}$ from the localization entropy of a flat distribution.

A possible issue with calculating positional information within a coarse-grained model, is that the obtained value is an underestimate. This is the case if the localization is confined to a region approximately the size of a unit cell. Since we find the localizations of genomic regions to be significantly larger than this (Fig. 2B), we do not expect our estimate to be sensitive to the course graining scale.

## S11 Local extension interval and origin of *ori* and *ter* extensions

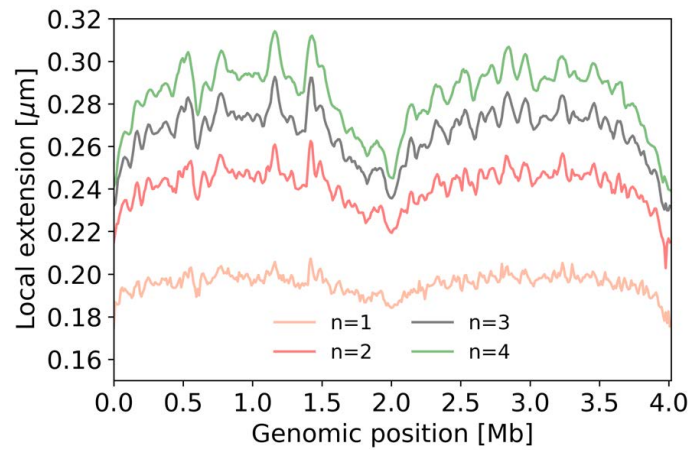

Figure S16: Local extensions, defined as the average distance between the $n^{\text{th}}$ nearest neighbours of a genomic region, shown for $n = 1$ up to $n = 4$. The value of $n = 2$ is shown in the main text as its features are more prominent than those for $n = 1$, but less smoothed out than for higher values of $n$. The locations of the peaks are largely identical between these different choices for $n$.

A possible explanation for the low local extension of the *ori* and *ter* regions, would be the turning around of the average long-axis positions at these regions. As the local extension of a region is calculated as the average geometric distance between its $n^{th}$ neighbours, such an effect could cause the observed low local extension. To test if this is the case, we make use of the presence of variations in the positions of the *ori* and *ter*; for a subset of states, these will not be the furthest

24

regions along the long axis. If the inferred low local extension is indeed due to a 'turning around' of the chromosome at the *ori* and *ter*, the local extension would be expected to be higher for this subset of states.

Taking a conditional average of the local extension of the *ori* over states where the previous 5 or subsequent 5 genomic regions all have an equal or lower long-axis position than the *ori* region, we find an increase of only 2% compared to an average over all states. For the *ter* region, we find the same statistics (2% increase if either set of 5 neighboring regions has a higher or equal long-axis position than the *ter*). Thus, the inferred local density of the *ori* and *ter* regions reflect the intrinsic extensions of these regions, rather than artefacts due to a turning around of the average long axis positions at these sites.

## S12  Independence of results for modified MaxEnt models

To test if our results are robust under minor model modifications, we inferred two alternative MaxEnt models: one with a slightly curved confinement, and one with a tethered *ori*. The former incorporates the typically observed *C. crescentus* cell shape, the latter enforces the experimentally measured long-axis distribution of the position of the *ori* locus. The inferred models are shown in Figs. SS18 and SS19.

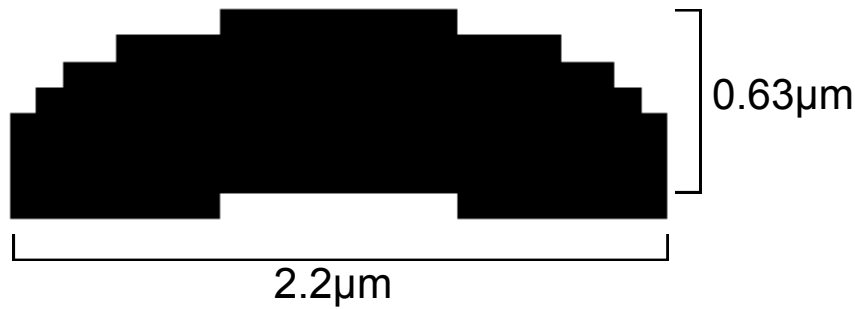

Figure S17: Top view of the curved cell shape used for analyses presented in this section.

25

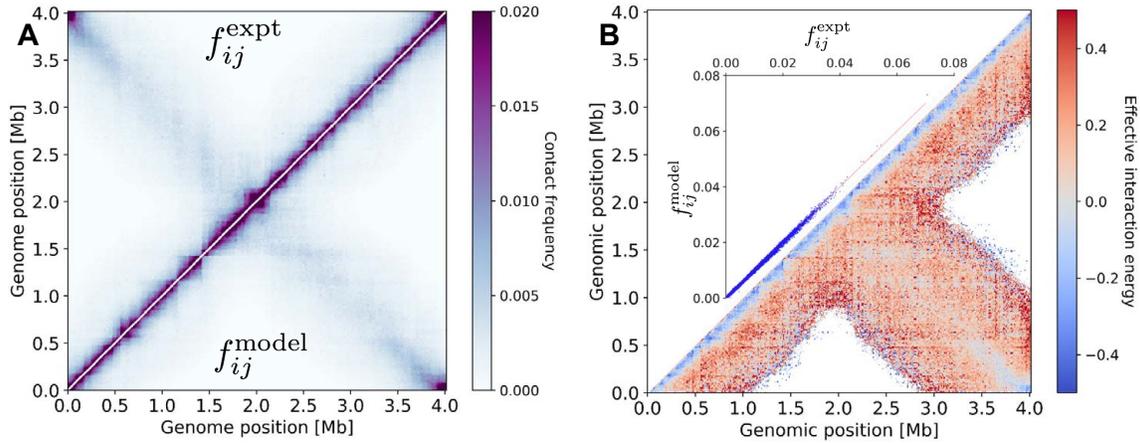

Figure S18: Results for main text Figure 1, re-analyzed for a model with tethered *ori*.

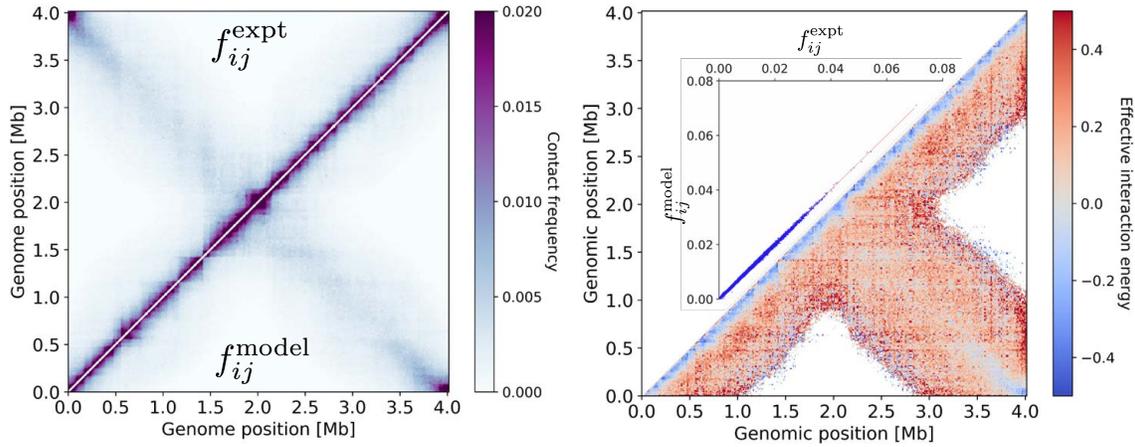

Figure S19: Results for main text Figure 1, re-analyzed for a model with a curved cell.

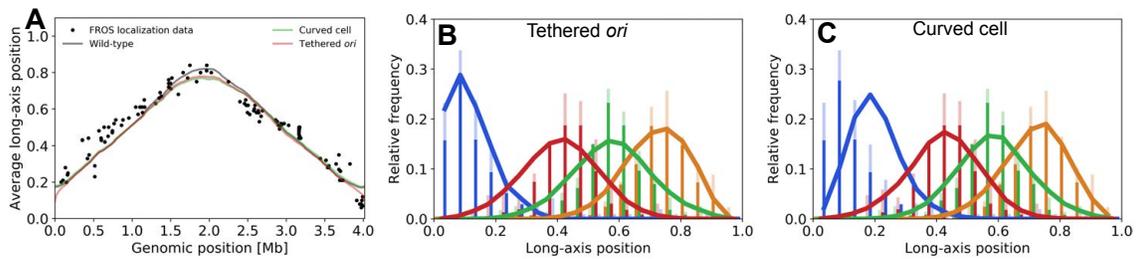

Figure S20: Results for main text Figure 2, re-analyzed for a model with a tethered *ori* and a curved cell.

26

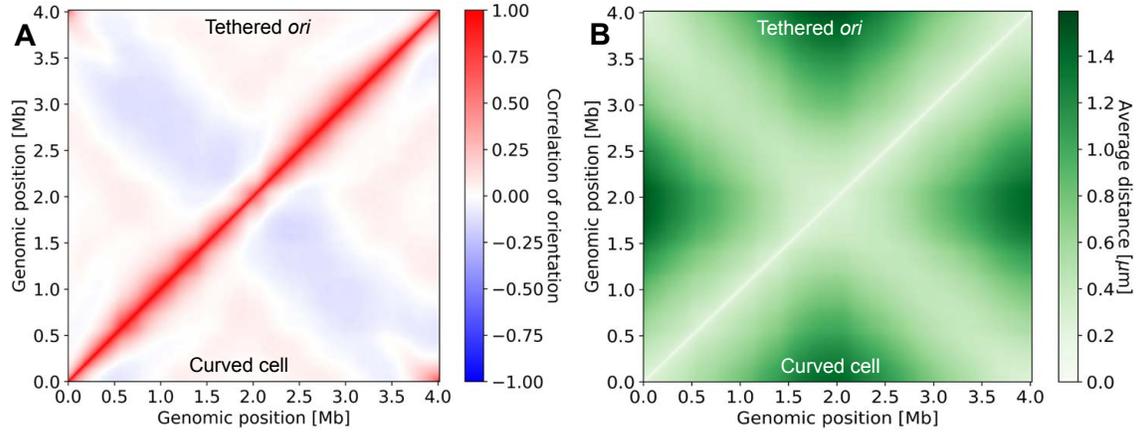

Figure S21: Results for main text Figure 3, re-analyzed for a model with a tethered *ori* and a curved cell.

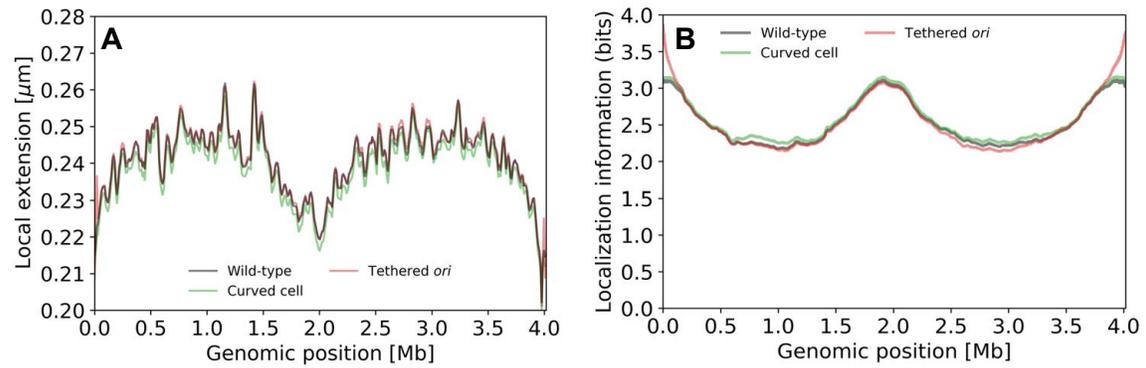

Figure S22: Results for main text Figure 4, re-analyzed for a model with a tethered *ori* and a curved cell.



\end{document}